\documentclass[twocolumn,trackchanges,openany]{aastex631}

\usepackage{natbib}
\usepackage{graphicx}
\usepackage{epstopdf}
\usepackage{amsmath}
\usepackage{multirow}   

\let\oldAA\AA
\renewcommand{\AA}{\text{\normalfont\oldAA}}

\newcommand{\fst}[1]{#1}
\newcommand{\fracA}{{\rm frac_{AGN}}}
\newcommand{\sersic}{S\'{e}rsic}

\begin{document}

\title{Analysis of Multi-epoch JWST Images of $\sim 300$ Little Red Dots: Tentative Detection of Variability in a Minority of Sources
}

\author[0000-0002-2420-5022]{Zijian Zhang}
\affiliation{Kavli Institute for Astronomy and Astrophysics, Peking University, Beijing 100871, China}
\affiliation{Department of Astronomy, School of Physics, Peking University, Beijing 100871, China}

\author[0000-0003-4176-6486]{Linhua Jiang}
\affiliation{Kavli Institute for Astronomy and Astrophysics, Peking University, Beijing 100871, China}
\affiliation{Department of Astronomy, School of Physics, Peking University, Beijing 100871, China}

\author[0000-0002-4385-0270]{Weiyang Liu}
\affiliation{Kavli Institute for Astronomy and Astrophysics, Peking University, Beijing 100871, China}
\affiliation{Department of Astronomy, School of Physics, Peking University, Beijing 100871, China}

\author[0000-0001-6947-5846]{Luis C. Ho}
\affiliation{Kavli Institute for Astronomy and Astrophysics, Peking University, Beijing 100871, China}
\affiliation{Department of Astronomy, School of Physics, Peking University, Beijing 100871, China}

\begin{abstract}
James Webb Space Telescope (JWST) has revealed a population of red and compact sources at $z \gtrsim 5$ known as ``Little Red Dots'' (LRDs) that are likely active galactic nuclei (AGNs). Here we present a comprehensive study of the variability of 314 LRDs with multi-epoch JWST observations in five deep fields: UDS, GOODS-S, GOODS-N, Abell 2744, and COSMOS. Our analyses use all publicly available JWST NIRCam imaging data in these fields, together with multi-epoch JWST MIRI images available. We measure the significance (signal-to-noise ratio or ${\rm SNR}_{\rm var}$) of the variabilities for all LRDs and statistically evaluate their variabilities using the ${\rm SNR}_{\rm var}$ distributions. We pay particular attention to the systematic offsets of photometric zero points among different epochs that seem to commonly exist. The derived ${\rm SNR}_{\rm var}$ distributions of the LRDs, including those with broad H$\alpha$/H$\beta$ emission lines, follow the standard Gaussian distribution, and are generally consistent with those of the comparison samples of objects detected in the same images. This finding suggests that the LRD population on average does not show strong variability, which can be explained by super-Eddington accretion of the black holes in AGNs. Alternatively, many of them may be dominated by galaxies. We also find eight strongly variable LRD candidates with variability amplitudes of 0.24 -- 0.82 mag. The rest-frame optical SEDs of these variable LRDs should have significant AGN contribution. Future JWST observations will provide more variability information of LRDs.

\end{abstract}

\section{Introduction}
\label{sec:intro}
James Webb Space Telescope is now revolutionizing our understanding of the early Universe. Among its discoveries, the identification of a class of red and compact objects has been particularly intriguing. These objects exhibit a unique ``v'' shape spectral energy distribution (SED) that is red in the rest-frame optical and blue in the rest-frame ultraviolet (UV). Their point-like morphology and red colors in the $\sim 2$--$5 ~\mu m$ range (observed-frame) have earned them the name ``Little Red Dots'' \citep[LRDs; e.g.,][]{2023Natur.616..266L,2024ApJ...963..128B,2024ApJ...963..129M}. LRDs are found to be ubiquitous from redshifts $z \sim 4$ up to $z \sim 9$ \citep[e.g.,][]{2023ApJ...954L..46L,2024ApJ...963..128B,2024arXiv240403576K,2024ApJ...968...38K}. Up to now, spectroscopy of more than 60 photometrically identified LRDs reveal that $\gtrsim 70\%$ of them exhibit broad Balmer lines \citep[e.g.,][]{2023ApJ...952..142F,2023ApJ...959...39H,2024ApJ...964...39G,2024arXiv240618207I,2024A&A...691A..52K,2024ApJ...963..129M,2024arXiv240302304W,2024ApJ...968...34W}. Despite these observations, the true nature of LRDs remains highly debated.

The rest-frame UV-optical SEDs of these objects were initially interpreted as galaxies with high stellar masses ($\gtrsim 10^{10} ~\rm  M_{\odot}$), and their red colors are due to strong Balmer breaks or dusty star formation \citep[e.g.,][]{2023Natur.616..266L,2023ApJ...956...61A}. JWST/MIRI observations at longer wavelength reveal a notably flat SED in the rest-frame \hbox{mid-infrared} (mid-IR) for several LRDs, which also suggests the presence of a 1.6 $\rm \mu  m$ stellar bump characteristic of stellar emission \citep[e.g.,][]{2024ApJ...968....4P,2024ApJ...968...34W}. Such galaxy-only models are further supported by the X-ray weakness of most LRDs \citep[e.g.,][]{2024arXiv240610341A,2024ApJ...969L..18A,2024arXiv240704777K,2024arXiv240500504M,2024ApJ...974L..26Y}, and the prominent Balmer breaks observed in several LRDs \citep[e.g.,][]{2023ApJ...955L..12B,2024ApJ...975..178K,2024arXiv240302304W}. The inferred high masses with small effective radii ($\sim 100$ pc) would indicate extremely high stellar mass densities \citep[e.g.,][]{2023ApJ...955L..12B,2024RNAAS...8..207G}, raising the possibility that the broad lines may originate from the galaxies' kinematics.

A more natural explanation of the broad lines and compact sizes of LRDs is that they are active galactic nuclei (AGNs). \citet{2024ApJ...964...39G} explained the SED of LRDs using a scattered AGN component in the UV combined with an intrinsically reddened AGN component in the optical. Recently, \citet{2025ApJ...980...36L} suggested that LRDs can be explained by a typical AGN SED reddened by a UV-flattened extinction law similar to that seen in dense regions like the Orion Nebula or certain AGN environments. They also
demonstrated that the flat mid-IR SED of LRDs may arise from an extended dust and gas distribution in the torus. In the AGN-only scenario, the observed Balmer break of some LRDs can be caused by extremely dense gas in the line of sight, and such dense gas can also naturally explain the presence of Balmer absorption features in many high-redshift AGNs \citep{2025ApJ...980L..27I,2025arXiv250113082J}. Additionally, the observed X-ray weakness of LRDs could be due to super-Eddington accretion of black holes (BHs) and does not contradict the AGN scenario \citep[e.g.,][]{2024arXiv241203653I,2024ApJ...976L..24M}.

Considering the co-existence of AGN and galaxy signature, many works invoke AGN-galaxy hybrid models to explain the SED of LRDs. For example, some SED fitting of the photometric SED and/or NIRSpec/PRISM data for LRDs favors a massive galaxy dominating the rest-frame UV alongside a dust-reddened AGN contributing the rest-frame optical emission \citep[e.g.,][]{2024A&A...691A..52K,2024arXiv241006257M,2024arXiv240302304W,2024ApJ...969L..13W}. The emission line diagnostics for an unresolved LRD by \citet{2025ApJ...980L..29A} also suggested that star-forming galaxy component dominates in the rest-frame UV, while AGN dominates in the rest-frame optical.

Given these ambiguous results, it becomes evident that relying solely on static SEDs/spectra may not suffice to determine the nature of LRDs, and alternative approaches are needed. One promising method is to investigate the variability of LRDs. AGNs generally exhibit distinctive variability compared with pure galaxies \citep[e.g.,][]{1997ARA&A..35..445U,2004ApJ...601..692V,2007AJ....134.2236S}, which may be related to the accretion disk instabilities \citep[e.g.,][]{1997ARA&A..35..445U}. The AGN variability amplitude depends on the wavelength, luminosity, Eddington ratio $\lambda_{\rm Edd}$, BH mass, etc \citep[e.g.,][]{2004ApJ...601..692V,2011ApJ...728...26M,2012ApJ...758..104Z}. Assuming an Eddington ratio of 0.1, an AGN with a BH mass similar to those inferred from typical LRDs (e.g., $M_{\rm BH} \sim 10^{7}M_{\odot}$) is expected to show variability of $\gtrsim 0.1$ mag on the time scale of a few months according to the empirical model in \citet{2023MNRAS.518.1880B}.

The unprecedented sensitivity of JWST has already enabled the discovery of extremely faint transient and variable sources at high redshifts \citep[e.g.,][]{2023ApJS..269...43Y, 2025ApJ...979..250D}. Up to now, some observations in specific JWST deep fields have spanned more than two years, corresponding to approximately three months in the rest frame at $z = 7$. This time baseline is sufficient to probe the variability of low-mass AGNs, such as LRDs. \citet{2024arXiv240704777K} analyzed multi-wavelength multi-epoch NIRCam data for 3 LRDs and two broad H$\alpha$ emitters in Abell 2744 field and found no significant variability. Their sample size is small, and the non-detection of variability may be due to the limited sampling of the data.

In this work, we present a systematic investigation of the variability of LRDs based on a much larger sample and complete data in five JWST deep fields: UDS, GOODS-S, GOODS-N, Abell 2744, and COSMOS. We compile all publicly available JWST/NIRCam data and incorporate multi-epoch MIRI data for the first time. We apply careful photometric zero point offset correction and uncertainty calibration to ensure reliable results. 
The paper is organized as follows. In Section \ref{sec:sample}, we introduce our LRD sample. In Section \ref{sec:variability_analyses}, we describe our data and data processing, the correction of systematic photometric zero point offsets, and the calibration of photometric uncertainties. The results of the variability analyses of the whole LRD sample and several variable LRD candidates are shown in Section \ref{sec:results}. In Section \ref{sec:discussion}, we discuss the effect of the variation of the point spread function (PSF) and the indications of our results. Conclusions are given in Section \ref{sec:conclusion}. Throughout this paper, we use a cosmology with $H_{0}=67.4~ \rm km~s^{-1}~Mpc^{-1}$, $\Omega_{\rm M} = 0.315$, and $\Omega_{\rm \Lambda} = 0.686$ \citep{2020A&A...641A...6P}. By default, measurement uncertainties are quoted at a 1$\sigma$ confidence level, while upper limits are quoted at a 90\% confidence level.

\section{Samples of LRDs}
\label{sec:sample}
We collected LRD samples from the literature. The main samples are from \citet{2024arXiv240610341A}, \citet{2024arXiv240403576K}, and \citet{2024ApJ...968...38K}, which all present large samples of LRDs.

The LRDs in \citet{2024ApJ...968...38K} were selected from several deep JWST/NIRCam fields totaling $\sim 340~\rm arcmin^2$, including CEERS \citep[PID1345;][]{2023ApJ...946L..12B}, PRIMER in COSMOS and UDS (PID1837), JADES \citep[PID1180, 1210, 1286, 1287;][]{2023arXiv230602465E,2023arXiv231012340E} and FRESCO \citep[PID1895;][]{2023MNRAS.525.2864O} in GOODS-S and JEMS \citep[PID1963;][]{2023ApJS..268...64W}. The sample selection of \citet{2024ApJ...968...38K} follows the method in \citet{2025ApJ...978...92L}. For low redshift ($z<6$) LRDs, they used a color cut of $\rm F115W - F150W < 0.8$ at $1~\mu \rm m < \lambda < 2~\mu \rm m$ and two color cuts of  $\rm F200W - F277W > 0.7$ and $\rm F200W - F356W > 1.0$  at $\lambda > 2~\mu \rm m$. 
For higher-redshift LRDs, they used the color cuts of $\rm F277W - F356W > 0.6$, $\rm F277W - F444W > 0.7$, and $\rm F150W - F200W < 0.8$. Sources were required to be spatially unresolved in F444W with $f_{\rm F444W}({0\farcs{4}})/f_{\rm F444W}(0\farcs{2}) < 1.7$ and removed the contamination of brown dwarfs using F115W $-$ F200W $ > -0.5$ as suggested by \citet{2025ApJ...978...92L}.

The 341 LRDs presented in \citet{2024arXiv240403576K} are from the CEERS, PRIMER, JADES, UNCOVER \citep[PID2561;][]{2024ApJ...974...92B} and NGDEEP \citep[PID2079;][]{2023ApJ...954...31C} Surveys. 
They selected red sources with a UV excess using an optical continuum slope cut of $\beta_{\rm opt} > 0$ and a UV slope cut of $\beta_{\rm UV} < -0.37$. These sources were required to be compact with a half-light radius smaller than 1.5 times that of a star in the F444W band. Furthermore, they used additional slope cuts to remove contamination from brown dwarfs ($\beta_{\rm UV} < -2.8$) and $\beta_{\rm opt}$ boosted due to strong line emission affecting one or more bands ($\beta_{\rm F277W-F356W} > -1$ at $z < 8$ and $\beta_{\rm F277W-F410M} > -1$ if F410M is available). Their method can select LRDs over a wide redshift range ($z \sim $ 2--11) and is less susceptible to contamination from galaxies with strong breaks.

\citet{2024arXiv240610341A} selected 434 LRDs from the 0.54 $\rm deg^2$ COSMOS-Web Survey \citep[PID1727;][]{2023ApJ...954...31C}. They selected compact sources brighter than 27.5 mag in F444W, but did not apply a blue rest-frame UV criterion, which is different from \citet{2024arXiv240403576K} and \citet{2024ApJ...968...38K}. Instead, they focused on the red end of the spectrum with F277W $-$ F444W $> 1.5$, which is a more extreme red color selection compared to other studies. This color cut biases the sample towards $z \geq 5$. Focusing on the reddest objects largely mitigates the contamination from extreme emission line galaxies. Their sample includes 37 LRDs covered by the PRIMER survey, and $\sim 90\%$ of these satisfy the multi-band color selection criteria described in \citet{2025ApJ...978...92L}.

\begin{table}[t!]
\begin{center}
   \caption{Number of LRDs in each field.}
   \label{tab:num_of_sample}
   \hspace{-2pt}
\begin{tabular}{llll}
\hline\hline
Field & Total   &With multi-epoch   & Broad-line \\
&sample& observations & sample   \\[4pt]
\hline
UDS  & 165 & 97 &6 \\
GOODS-S & 84 & 70& 1         \\
GOODS-N & 8& 8  & 8      \\
Abell 2744 & 29 & 25 & 15        \\
COSMOS & 520&114& 1   \\
\hline
Total number& 806 & 314 & 31  \\
\hline 
   \end{tabular}
\end{center}
\end{table}

\begin{figure*}
 \includegraphics[width=0.95\textwidth]{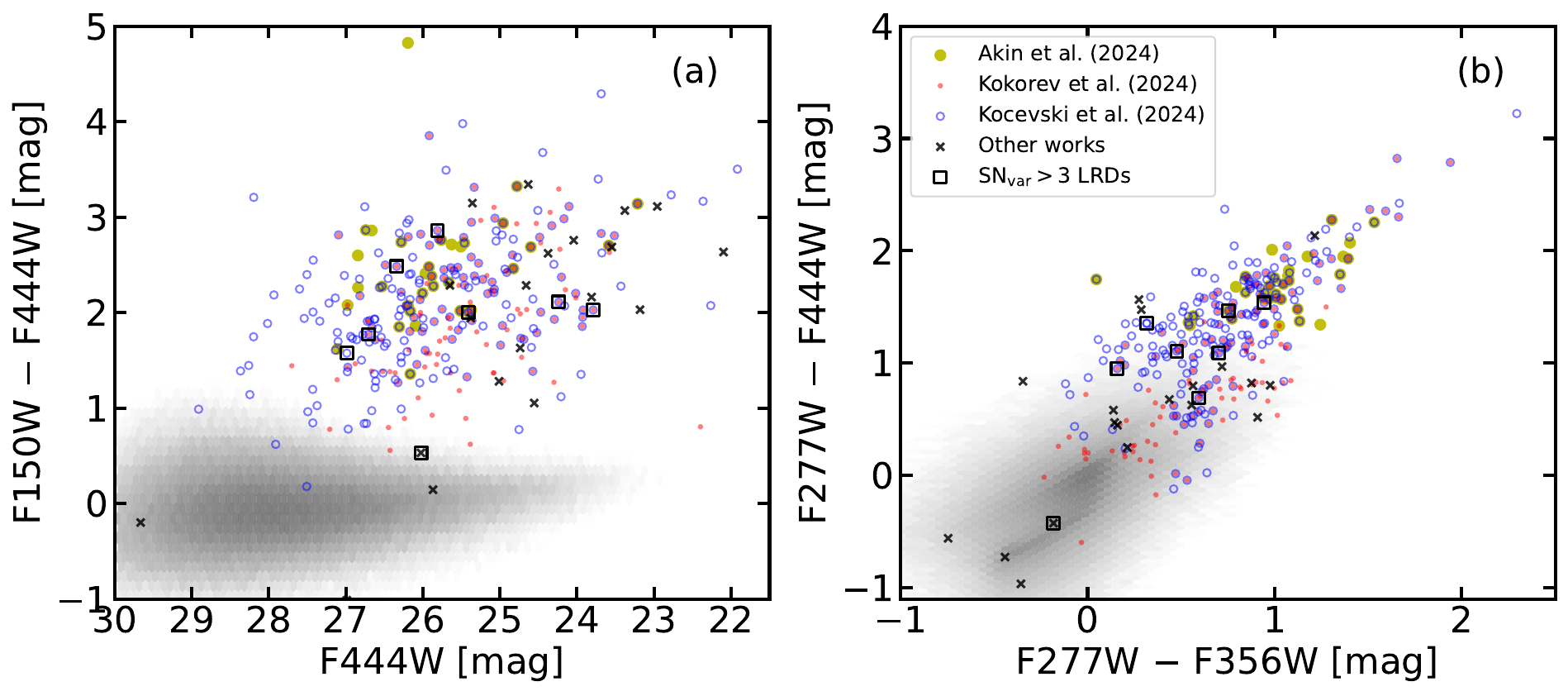}
 \centering
 \caption{(a) Color-magnitude diagram of F444W vs. F150W$-$F444W and (b) color-color diagram of F277W$-$F356W vs. F277W$-$F444W for the 314 LRDs that have multi-epoch detections. The LRDs from \citet{2024arXiv240610341A}, \citet{2024arXiv240403576K}, and \citet{2024ApJ...968...38K} are denoted as the large solid yellow dots, blue open circles, and small solid red dots, respectively. LRDs from other works are denoted as the black crosses. The eight variable LRD candidates in Section \ref{subsec:individual_LRD_variability} are enclosed in the black squares. For comparison, all sources identified in these five fields are shown in gray.}
 \label{fig:color_color}
\end{figure*}

We merged these three samples and removed duplicate sources based on their coordinates. Any two sources with a separation smaller than $1''$ were treated as the same source. The merged sample consist of 853 different LRDs. We considered additional LRD samples from other works \citep{2023ApJ...952..142F,2023ApJ...959...39H,2024ApJ...964...39G,2024A&A...691A..52K,2024ApJ...963..129M,2024arXiv240302304W,2024ApJ...968...34W}. We included eight broad H$\alpha$ emitters that have extended morphologies and flatter rest-frame optical SEDs than LRDs in \citet{2023ApJ...959...39H} and an extended source at $z = 6.6$ (\textit{Virgil}) detected by JWST/MIRI with photometric properties similar to those of LRDs \citep{2024arXiv240618207I},

Our final LRD sample consists of 907 LRDs in total. We give each LRD an I.D. (LID 1--907) in order of R.A. 806 LRDs are located in the UDS, GOODS-S, GOODS-N, Abell 2744, and COSMOS fields, forming a parent sample for our variability analyses. Table \ref{tab:num_of_sample} summarizes the number of LRDs in each field. Among this sample, 31 LRDs are spectroscopically confirmed to have broad Balmer lines. We adopted their BH mass measurement in the literature \citep{2023ApJ...959...39H,2024arXiv240610341A,2024ApJ...964...39G,2024A&A...691A..52K,2024arXiv240403576K,2024ApJ...975..178K,2024arXiv240500504M,2024ApJ...963..129M,2024arXiv240302304W,2024ApJ...968...34W}. 
In this sample, 314 LRDs have multi-epoch observations that allow us to study their variabilities. Their color-magnitude diagram (F444W v.s. F150W$-$F444W) and color-color diagram (F277W$-$F356W v.s. F277W$-$F444W) are shown in Figure \ref{fig:color_color}. LRDs from \citet{2024arXiv240610341A} are significantly redder than those from other works, while LRDs from \citet{2024arXiv240403576K} and \citet{2024ApJ...968...38K} occupy larger color space.

\section{Data, Processing, and Calibration}
\label{sec:variability_analyses}
\subsection{NIRCam Data Processing}
\label{subsec:NIRCAMdata}

We download raw JWST/NIRCam image {\tt\string uncal.fits} files in the A2744, GOODS-S, GOODS-N, COSMOS, and UDS fields from the MAST archive. The detailed information of these files is summarized in Table \ref{table:nircam}. For the COSMOS field, we only reduce data in the PRIMER-COSMOS region since the other regions have few overlapping multi-epoch observations. We reduce the raw data using the combination of the \textit{JWST Calibration Pipeline} (v.1.12.5), the CEERS NIRCam imaging reduction,\footnote{\url{https://github.com/ceers/ceers-nircam.}} and our own custom codes. We use CRDS version 11.17.0 and CRDS context 1252. The details of the CEERS NIRCam imaging reduction are presented in \citet{2023ApJ...946L..12B}. Here, we provide a brief description of the steps.

Stage 1 of the JWST Calibration Pipeline performs detector-level corrections, starting from {\tt\string uncal.fits} files and ending with {\tt\string rate.fits} files in units of count/s. Following the CEERS team's scripts, before running \hbox{ramp-fitting} step in Stage 1 of the pipeline, we identify snowballs, increase their footprints, and flag them as {\tt\string JUMP\_DET} so that snowballs are efficiently removed after the ramp-fitting step. ``Wisp'' features from detector B4 of the F150W, F200W, and F210M images are subtracted using the wisp templates provided by the NIRCam team\footnote{\url{https://stsci.app.box.com/s/1bymvf1lkrqbdn9rnkluzqk30e8o2bne.}}. The 1/f noise, i.e., horizontal and vertical striping patterns in the images, are subtracted with {\tt\string remstriping.py} in the CEERS team's scripts.

\begin{table*}

\caption{Information of the NIRCam images.}
\hspace*{-1cm}
\begin{tabular}{c c c c c c} 

 \hline\hline
 Field & Filters & {\tt\string DATE-OBS} & Central R.A. & Central Decl. & PIDs \\ 
 \hline
 A2744     & F070W F090W F115W F140M & 2022-06-28 & $0^{\rm h}14^{\rm m}13^{\rm s}_{.}92$ &  $-30^{\rm d}25^{\rm m}04^{\rm s}_{.}80$ & 1324 2561 2756\\
           & F150W F162M F182M F200W & -- 2023-12-10   &    &    & 2883 3516 3538\\
           & F210M F250M F277W F300M &              &         &           & 3990 4111     \\
           & F335M F356W F360M F410M &              &         &           &               \\
           & F430M F444W F460M F480M &              &         &           &               \\
 \hline

 GOODS-S   & F090W F115W F150W F162M& 2022-09-29 & $3^{\rm h}32^{\rm m}35^{\rm s}_{.}16$ &  $-27^{\rm d}51^{\rm m}21^{\rm s}_{.}60$  & 1180 1210 1283\\
           &  F182M F200W F210M F250M& -- 2024-02-01   &    &    & 1286 1895 1963\\
           &  F277W F300M F335M F356W &              &          &           & 2079 2198 2514\\
           & F410M F444W       &              &          &           & 3215 3990 6541\\
 \hline

 GOODS-N   & F070W F090W F115W F150W & 2023-02-03  & $12^{\rm h}37^{\rm m}28^{\rm s}_{.}20$ &  $62^{\rm d}14^{\rm m}00^{\rm s}_{.}60$   & 1181 1895 2514\\
           & F182M F200W F210M F277W & -- 2024-05-19   &    &    & 2674 3577     \\
           & F335M F356W F410M F444W &              &           &          &               \\
 \hline

 COSMOS    & F090W F115W F150W F200W & 2022-11-25  & $10^{\rm h}00^{\rm m}30^{\rm s}_{.}12$ &  $2^{\rm d}19^{\rm m}48^{\rm s}_{.}00$ & 1635 1727 1810\\
           & F277W F356W F410M F444W & -- 2024-05-21   &    &    & 1840 1933 2321\\
           &                         &              &           &         & 2362 2514 3990\\
           &                         &              &           &         & 6585          \\
 \hline

 UDS       & F090W F115W F150W F200W & 2022-07-29 & $2^{\rm h}17^{\rm m}22^{\rm s}_{.}44$ &  $-5^{\rm d}13^{\rm m}19^{\rm s}_{.}20$ & 1837 1840 2514\\
           & F277W F356W F410M F444W & --2024-07-23    &    &    & 3990          \\
 \hline
\end{tabular}

\label{table:nircam}
\end{table*}

We then run Stage 2 of the JWST Calibration Pipeline with the default parameters, which involves individual image calibrations such as flat-fielding and flux calibration. The output files are {\tt\string cal.fits}. We group these files according to {\tt\string DATE-OBS} in their headers, and we run Stage 3 of the JWST Calibration Pipeline to obtain a single mosaic for each {\tt\string DATE-OBS} and filter in each field. Astrometry is calibrated using {\tt\string TweakregStep} with {\tt\string abs\_refcat} set to user-provided reference catalogs in corresponding fields. For A2744, we use the UNCOVER DR2 catalog as the reference catalog \citep{2024ApJS..270....7W}. For GOODS-S/GOODS-N, we generate a reference catalog as follows. We first align all LW F277W, F356W, F410M, and F444W images to detected objects in the Hubble Legacy Field GOODS-S and GOODS-N F814W image, and then combine all these LW images to make a mosaic image. We finally detect objects in the mosaic image as the reference catalog. We make reference catalog for COSMOS and UDS using the same method and their LW images are aligned to the HSC COSMOS/UDS $y$-band catalog \citep{2022PASJ...74..247A}. 

\begin{figure*}
 \includegraphics[width=0.98\textwidth]{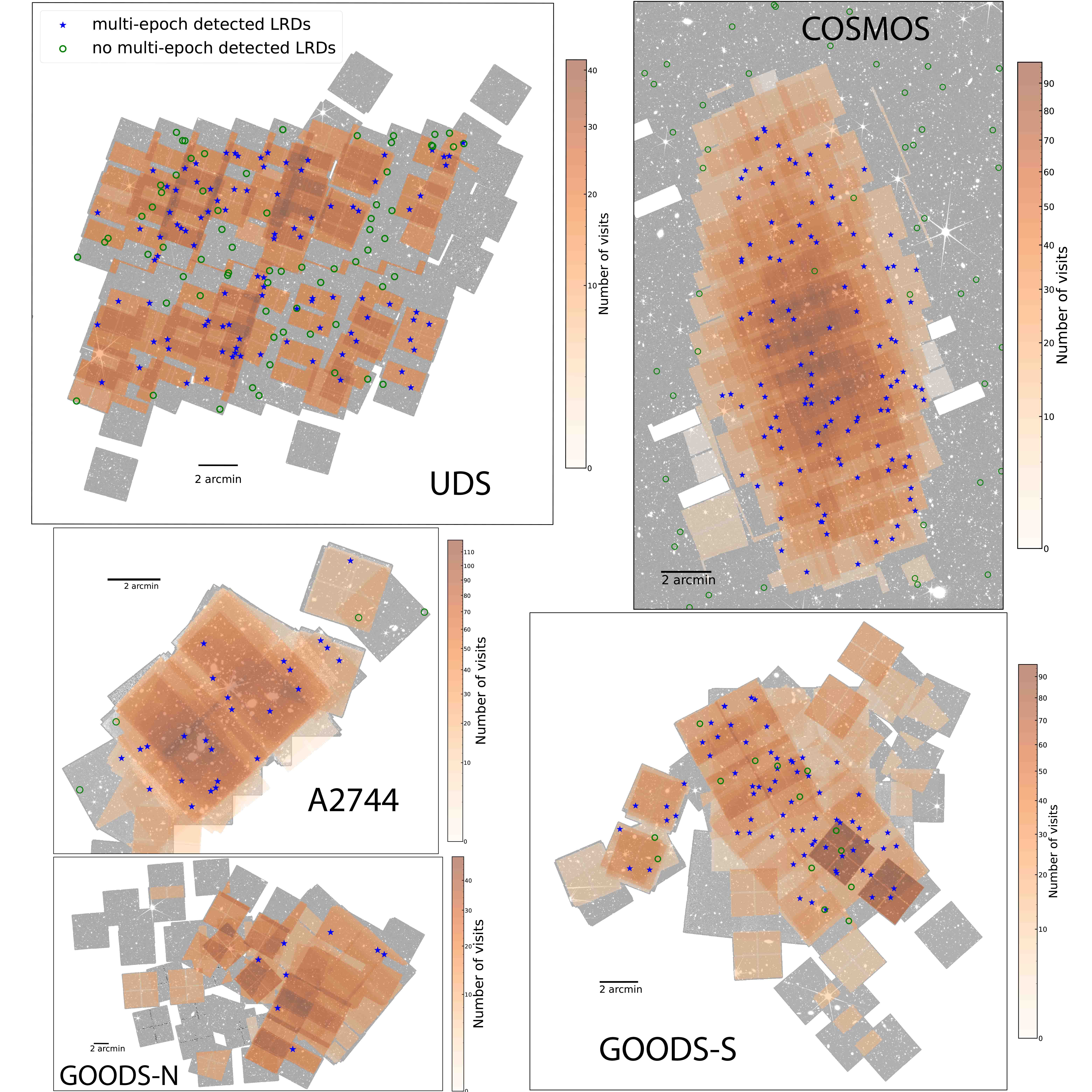}
 \centering
 \caption{JWST/NIRCam coverage of the five fields (north is up and east to the left) and the regions of multi-epoch NIRCam observations for UDS, A2744, GOODS-S, GOODS-N, and COSMOS fields. The background images are the co-added mosaic images of all bands in each field. Regions that overlap between \textit{visits} in the same band are highlighted in the orange shades. A darker color indicates more \textit{visits} with all bands combined, as shown in the color bars. The blue stars denote the LRDs with multi-epoch detections at least in one band. The green open circles denote the LRDs that do not have multi-epoch detections in any band.}
 \label{fig:mosaic_img}
\end{figure*}

After running {\tt\string SkyMatchStep} and {\tt\string OutlierDetectionStep}, we subtract a 2D background for each image following the method described in \citet{2023ApJ...946L..12B}, and then run {\tt\string ResampleStep} to drizzle and combine images to make one mosaic per {\tt\string DATE-OBS} per filter for each field. The data for one {\tt\string DATE-OBS} and one filter is defined as a \textit{visit}. We set {\tt\string pixfrac} to 0.8 and pixel scale to $0\farcs03$ for all filters. {\tt\string CRVAL} is the same for each filter, and {\tt\string CRPIX} is appropriately set so that all mosaics have the same WCS grid in each field. We also combine all images to make master mosaic images for each filter in each field. Finally, a 2D background is subtracted from each mosaic using the method described in \citet{2023ApJ...946L..12B}. Figure \ref{fig:mosaic_img} shows the NIRCam coverage of the five fields.

\subsection{MIRI Data Processing}
\label{subsec:MIRIdata}
We also make use of all public multi-epoch JWST/MIRI F560W and F770W imaging data in the GOODS-S field, collected from PID 1180, 1207, 1283, and 2516 \citep[e.g.,][]{2023arXiv231012340E,2024ApJ...975...83R}. We only consider F560W and F770W because LRDs are rarely detected in longer wavelengths, and these two bands have relatively large multi-epoch overlapping regions.

The MIRI data are reduced using v1.13.4 of the \textit{JWST Calibration Pipeline}, with CRDS version 11.17.14 and CRDS context 1215. Our reduction pipeline is based on the public pipeline from CEERS \citep{2023ApJ...956L..12Y},\footnote{\url{https://github.com/ceers/ceers-miri}.} but we add some additional custom steps to improve the quality of the data reduction. We follow the official pipeline in Stages 1 and 2 and remove the horizontal and vertical stripes after Stage 2. As described in \citet{2024ApJ...976..224A}, although the default pipeline contains two levels of outlier detection, faint outliers caused by warm pixels may escape both detection steps and result in noise peaks in the resampled image. Therefore, we median stack all the {\tt\string cal.fits} images after stripe removal (with pixels with DQ flag $>$ 4 masked) and construct a warm pixels map by finding pixels that are $> 3\sigma$ above the median of all pixels of the median-stacked image. Then, we manually mask these warm pixels in all {\tt\string cal.fits} images.

We also modify the super-background construction following the strategy explained in \citet{2023A&A...671A.105A} and \citet{2024ApJ...976..224A}. We first construct a reliable segmentation map by iteratively running the Stage 3 and masking the sources in each {\tt\string cal.fits} using the segmentation map of the mosaic image and median filtering out large gradients. A super background for a certain {\tt\string cal.fits} image is then constructed by median combining all other clean {\tt\string cal.fits}, with sources masked using the final segmentation map and subtracted. Then, we apply an additional $265\times265$ $\rm pix^2$ box median subtraction to remove any remaining varying background. For the background-subtracted {\tt\string cal.fits} files, we use the {\tt\string TweakregStep} in Stage 3 to perform astrometry calibration with a reference catalog based on the JWST/NIRCam F444W mosaic image. Finally, the astrometry-corrected and background-subtracted {\tt\string cal.fits} are processed through Stage 3, which produce the final mosaics with a pixel scale of 0\farcs{06}. For the MIRI data, we also produce the mosaic for each \textit{visit} and the master mosaic for each filter.

\subsection{Photometry}
\label{subsec:photometry}

For the image of each \textit{visit}, we perform aperture photometry using \texttt{SExtractor} \citep{1996A&AS..117..393B} to construct its source catalog. We first smooth the images with a $7\times 7$ Gaussian kernel with a FWHM of 3 pixels to detect faint sources. To minimize the detection of spurious sources, we mask 100 pixels along the detector borders for each image, as the borders usually have fewer exposures and result in excess noise and imperfect rejection of cosmic-ray signals. We use \texttt{SExtractor} in the dual image mode, and use the master mosaic image as the detection image. The \texttt{DETECT\_THRESH} parameter is set to be 3 and the \texttt{PHOT\_APERTURES} parameter is set to be $4\times{\rm FWHM}$ for each band. If $4\times{\rm FWHM}$ is smaller than 5 pixels, the aperture size is set to be 5 pixels. Table \ref{tab:sext_params} summarizes the \texttt{SExtractor} parameters used for photometry.

\begin{table}[t!]
\begin{center}
   \caption{Relevant \texttt{SExtractor} parameters for aperture photometry in Section \ref{subsec:photometry}.\label{tab:sext_params}}
   \hspace*{-0.5cm}
   \begin{tabular}{ll}
\hline\hline
Parameter                 & Value            \\[4pt]
\hline
\texttt{PHOT\_APERTURES}             & max($4\times$FWHM, 5 pixels)             \\
\texttt{DETECT\_MINAREA}  & 5                                    \\
\texttt{DETECT\_THRESH}   & 3                         \\
\texttt{ANALYSIS\_THRESH} & 3                             \\
\texttt{DEBLEND\_NTHRESH} & 32                   \\
\texttt{DEBLEND\_MINCONT} & 0.005                       \\
\texttt{FILTER\_NAME}     & gauss\_3.0                \\
\texttt{GAIN}             & 0                                 \\
\texttt{BACK\_TYPE}       & MANUAL                       \\
\texttt{BACK\_VALUE}      & 0                         \\
\texttt{BACK\_SIZE}       & 100                       \\
\texttt{BACK\_FILTERSIZE} & 3                         \\
\texttt{BACKPHOTO\_THICK} & 100                       \\
\hline
   \end{tabular}
\end{center}
\begin{minipage}{0.485\textwidth}{\small}

\end{minipage}
\end{table}

We find unusual detector artifacts in the NIRCam images of the COSMOS field observed on 2023-05-26 and the data of the GOODS-N field observed on 2023-05-28. These artifacts are bright and occupy large regions. They are likely caused by stray light. To address the issue of these contaminations, we mask the affected pixels using the following procedures. First, we smooth the image with a $5\times5$ top-hat PSF with a diameter of 5 pixels. We then run \texttt{SExtractor} with a \texttt{DETECT\_THRESH} parameter of 0.2, a  \texttt{DEBLEND\_MINCONT} parameter of 1, and a \texttt{DETECT\_MINAREA} parameter of 80,000. These parameters have been fine-tuned to reliably detect detector artifacts without detecting real sources of interest. The obtained segmentation maps are then used to mask the contamination regions. We also apply the same procedures to all other images to mask very large and bright sources.

For each \textit{visit} pair in each common band, we match their catalogs and build a \textit{visit}-pair catalog. We use the astroML \citep{astroML} \texttt{crossmatch\_angular} function to cross-match the object from one \textit{visit} catalog to another. As the \textit{visit} catalogs are created using the same detection image (cut-out images from the master mosaic image), the coordinates for the same sources in different \textit{visit} catalogs should be consistent. We use a maximum distance threshold of 0\farcs{02} to avoid mismatch. In each \textit{visit}-pair catalog, we calculate the magnitude difference $\Delta m = m_{\rm visit 1} - m_{\rm visit 2}$, the measured error of the magnitude difference $\Delta m_{\rm err} = \sqrt{m_{\rm visit 1,err}^2 + m_{\rm visit 2, err}^2}$, and the mean magnitude $\left \langle\Delta m_{\rm err}\right \rangle = (m_{\rm visit 1} + m_{\rm visit 2})/2$ for each sources. Figure \ref{fig:mosaic_img} shows the regions of multi-epoch NIRCam observations for the five fields. Some LRDs are located in regions that have multi-epoch observations but still do not have multi-epoch detections since they are too faint or masked out.

We consider a source as a variable source if it varied by more than 3 times the combined photometric uncertainty. Before we do this, we carefully examine photometric zero points of each individual \textit{visit} and photometric errors of sources. They are both are critical for the identification of variable sources.

\subsection{Correction of Systematic Photometric Zero Point Offset}
\label{subsec:correction_of_phot_offset}

\begin{figure}
 \includegraphics[width=0.48\textwidth]{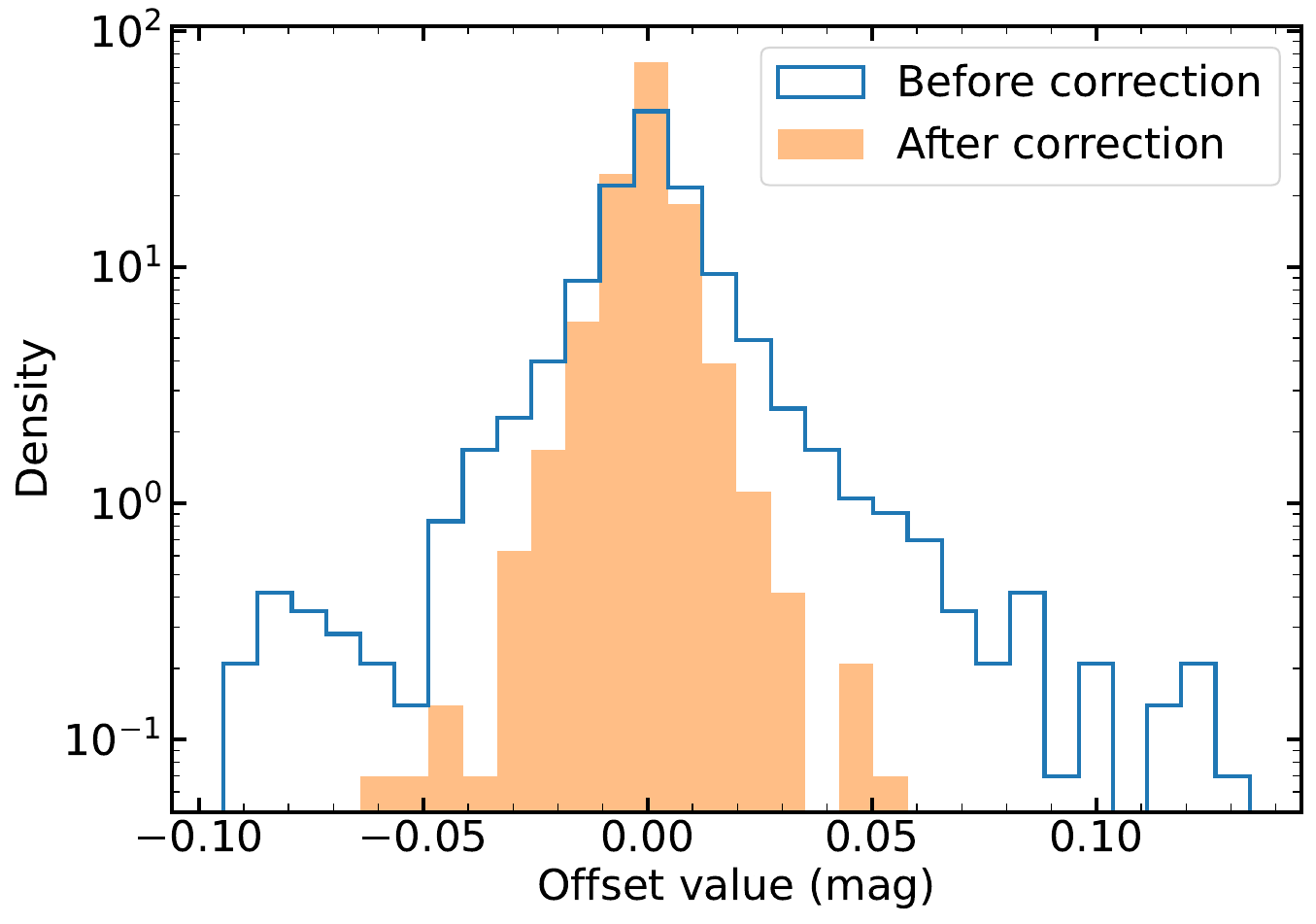}
 \centering
 \caption{Density distribution of systematic photometric zero point offsets of all NIRCam \textit{visit} pairs that have more than 50 sources before (blue solid line histogram) and after correction (orange shaded histogram).}
 \label{fig:phot_offset}
\end{figure}

For each \textit{visit} pair, we calculate its systematic photometric zero point offset as the median value of the iterative 3$\sigma$ clipped mean $\Delta m$ of sources with \texttt{SExtractor} \texttt{FLAGS} $< 1$ and \hbox{signal-to-noise} ratio (SNR) $>5$ in different magnitude bins. We verify that the offset is independent of magnitude for a certain \textit{visit} pair. Most of the \textit{visit} pairs show negligible offset, which confirms the photometric stability of NIRCam \citep{2023PASP..135d8001R}. However, some \textit{visit} pairs have large offsets of $\sim 0.1$ mag (see Figure \ref{fig:phot_offset}), which may be caused by the fluctuation of photometric zero point for different \textit{visit} and will bias the selection of variable sources. 

\begin{figure*}
 \includegraphics[width=0.95\textwidth]{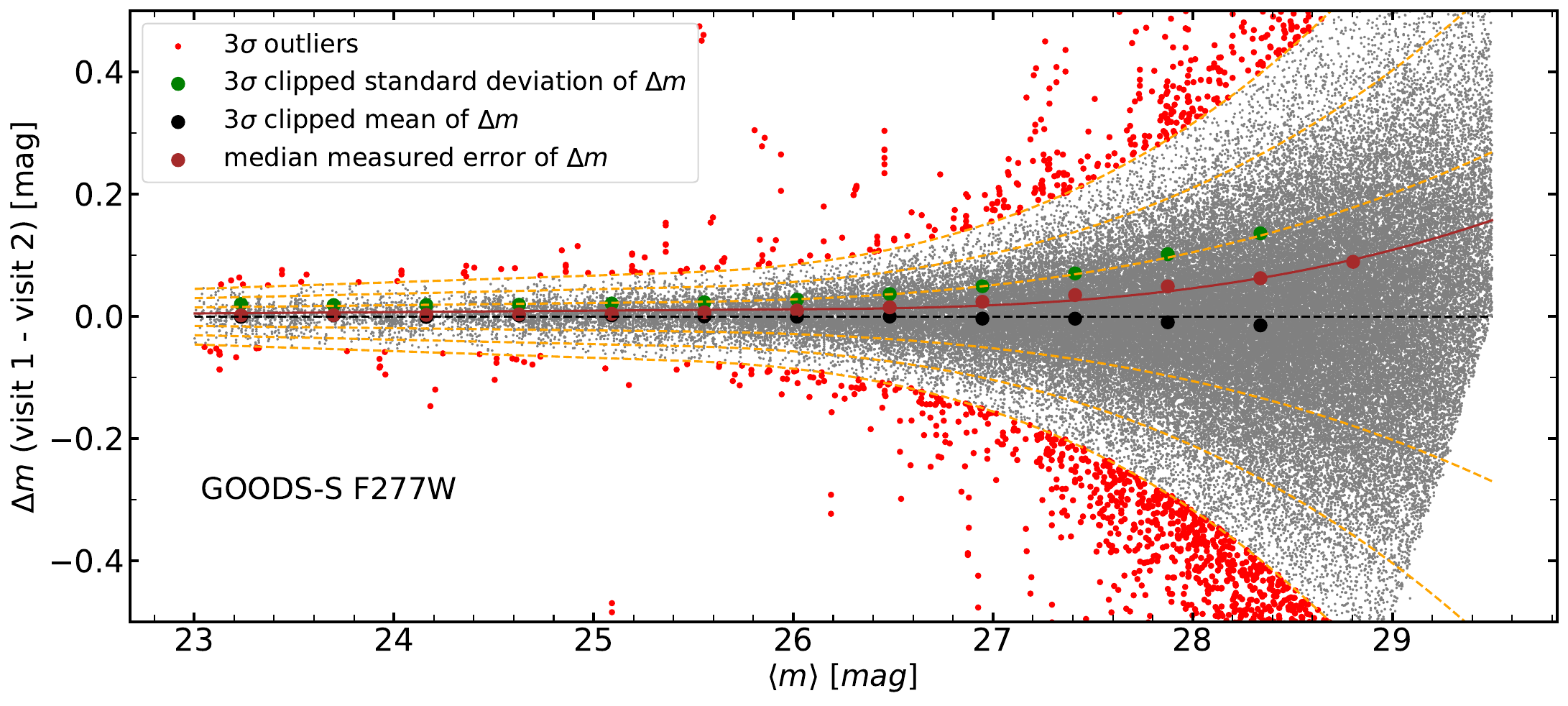}
 \centering
 \caption{Magnitude difference versus the mean magnitude of the F277W band for the GOODS-S field. The small gray dots are the $3\sigma$ clipped sample used to calculate the mean and standard deviation of $\Delta m$ in the magnitude bins, while the small red dots are clipped $3\sigma$ outliers. The green and black dots are the mean and standard deviation of $\Delta m$ in each bin, respectively. The brown dots are the median values of the measured errors of $\Delta m$ propagated from the errors of $m_{\rm visit1}$ and $m_{\rm visit2}$. The yellow dashed curves are the $1\sigma$, $2\sigma$, and $3\sigma$ deviation from the mean value. The black dashed curves represent $\Delta m = 0$.}
 \label{fig:scatter_example}
\end{figure*}

We correct the systematic photometric zero point offsets iteratively to achieve self-consistent results. We first select \textit{visit} pairs with more than 50 sources to ensure the offsets are reliably calculated. The zero point offset of each \textit{visit} is initially set as 0. For each pair, we apply the offset correction using the zero point offset difference between the two dates involved in the pair (they are all 0 in the first iteration). We then apply an iterative correction process to refine these offset correction values. In each iteration, the zero point offset of a certain \textit{visit} is adjusted by subtracting a correction value that is computed by averaging the offsets after applying the correction across all pairs involving this \textit{visit} date. This process is iterate for 500 times, with the offsets across all \textit{visit} pairs gradually converging towards a self-consistent result. The final corrections are applied to the original data and they significantly reduce the systematic photometric zero point offsets across all pairs. Figure \ref{fig:phot_offset} shows the density distribution of systematic photometric zero point offsets of all NIRCam \textit{visit} pairs before and after correction. Our correction largely reduces the offsets so that the corrected offsets of most \textit{visit} pairs are smaller than 0.02 mag. This step does not accurately calibrate the absolute photometric zero point of each \textit{visit}. Instead, it ensures that their relative photometric zero point is consistent, which is crucial in the variability analyses.

\subsection{Calibration of Photometric Uncertainties}
\label{subsec:correction_phot_uncertainty}

We calibrate the photometric uncertainties based on the variable source detection method described in previous works \citep[e.g.,][]{2006ApJ...639..731C,2020ApJ...894...24K,2024ApJS..272...19O}. This method assumes that most of the sources that we observed are not variable. Thus, the scatter of the magnitude difference for a large sample of sources reflects the real photometric uncertainty. To minimize the uncertainty introduced by saturated pixels, truncated footprints, corrupted apertures/footprints, or other issues, we ignore objects with \texttt{SExtractor} \texttt{FLAGS} $>$ 1. We also ignore objects with SNR $<5$ and with bad pixels (\texttt{IMAFLAGS\_ISO} $\neq 0$). Lastly, as most of the LRDs are point-like sources, we restrict the sources to have \texttt{ELONGATION} $<$ 1.5 to avoid the uncertainty introduced by the difference in morphology.

For each band in each common field, we plot the magnitude difference versus the mean magnitude combining all the \textit{visit}-pair catalogs. As an example, Figure \ref{fig:scatter_example} displays the results for the F277W band in the GOODS-S field. We divide the data into magnitude bins with a bin size of $\sim 0.5$ mag and compute the 3$\sigma$ clipped standard deviation of $\Delta m$ ($\sigma_{\Delta m}$) and the median of $\Delta m_{\rm err}$ ($\left \langle \Delta m_{\rm err} \right \rangle$) in each magnitude bin. The $\sigma_{\Delta m}$ and $\left \langle \Delta m_{\rm err}\right \rangle$ are both fitted with a straight line in the bright end and with a third-order polynomial line in the faint end. The two lines are smoothly connected. The error correction curve is derived as the difference between the the $\sigma_{\Delta m}$ curve and the $\left \langle \Delta m_{\rm err} \right \rangle$ curve.

When deriving the error correction curves, we also remove the LRDs and known AGNs in each field to avoid the influence of any variability of these sources. We collect AGN samples from the literature, and the details of the known AGN samples for each field are listed in Appendix (Table \ref{tab:known_AGNs}). For Abell 2744, there is no available AGN catalogs, and the fraction of typical AGNs should be relatively small in such a galaxy cluster field. Thus, we do not remove AGNs in this field.

\begin{figure}[ht]
 \includegraphics[width=0.45\textwidth]{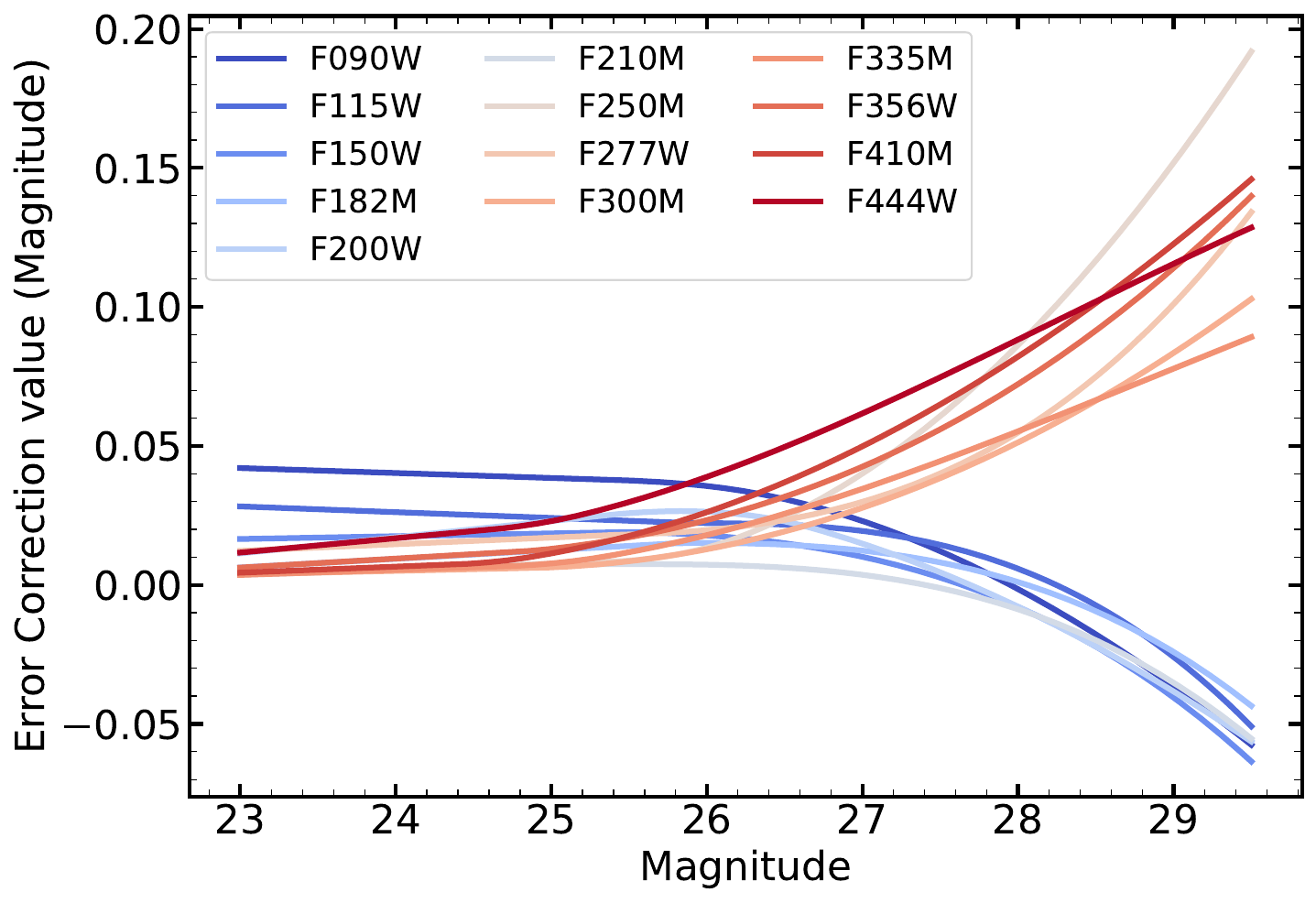}
 \centering
 \caption{The $\Delta m$ error correction curves of different bands for the GOODS-S field.}
 \label{fig:correction_curves}
\end{figure}

Figure \ref{fig:correction_curves} shows the correction curves for different bands of the GOODS-S field as an example. The correction curves for the LW and SW bands show different patterns. For the LW bands, the correction increases with magnitude, while for the SW bands the correction decreases with magnitude and even becomes negative in the bright end. We calibrate the error of $\Delta m$ and $m$ using the correction curves and the correction curves divided by the square root of two, respectively. We do the correction for each field separately, as we find that the correction curves also vary from field to field, likely caused by the different observational depths and dates. Lastly, we also set the minimum error for $m$ and $\Delta m$ to be 0.05 mag and 0.07 mag, respectively.

\section{Results}
\label{sec:results}

\subsection{No Detection of Significant Variability for the Whole LRD Sample}
\label{subsec:variability_of_LRD}

\begin{figure}[t]
\hspace{-0.4cm}
 \includegraphics[width=0.45\textwidth]{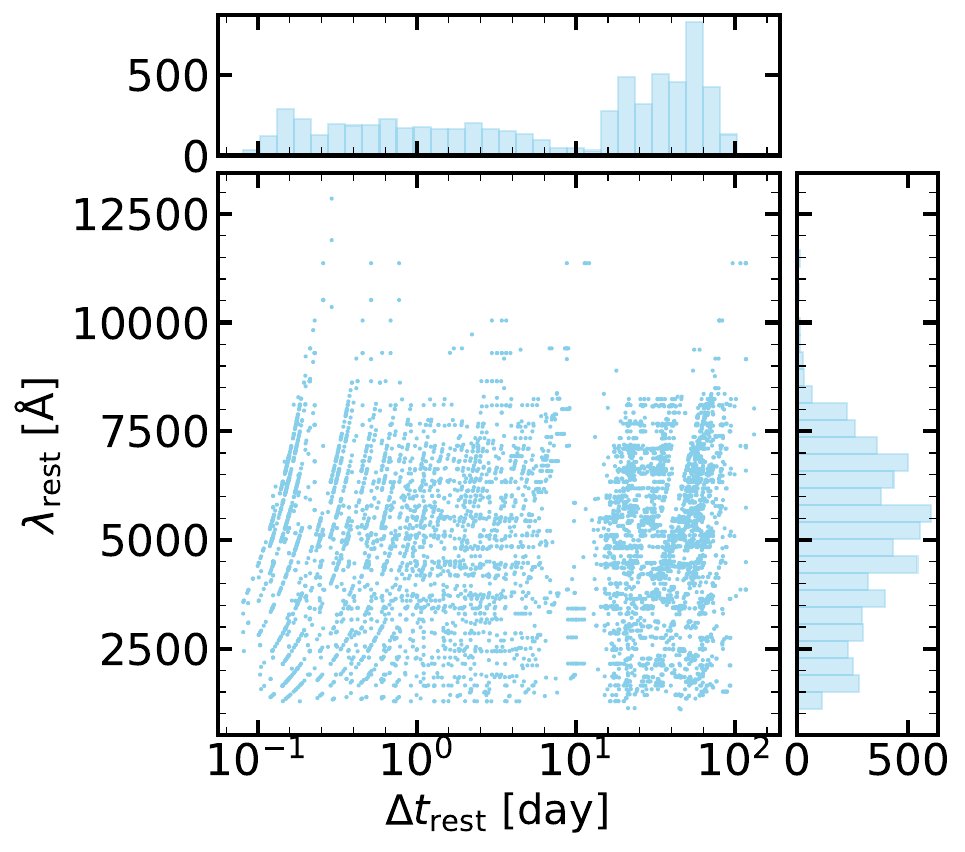}
 \centering
 \caption{Data points of the rest-frame time intervals and rest-frame wavelengths for the LRD sample. The histogram for each parameter is shown on each side.}
 \label{fig:restframe_wav_time}
\end{figure}

Our data probe the variability of the LRDs from rest-frame $\sim$1500 to $\sim$8500 \AA, in the rest-frame time scale of $\sim 0.1$ to $\sim 100$ days. Figure \ref{fig:restframe_wav_time} illustrates the distribution of the data points in terms of the rest-frame wavelength and time interval. We tend to have more data points in the rest-frame optical than those in the rest-frame UV, since LRDs are fainter in then rest-frame UV and thus are more difficult to have high SNR detectiosn on a \textit{visit} mosaic.

\begin{figure*}[ht]
\hspace{-0.4cm}
 \includegraphics[width=0.8\textwidth]{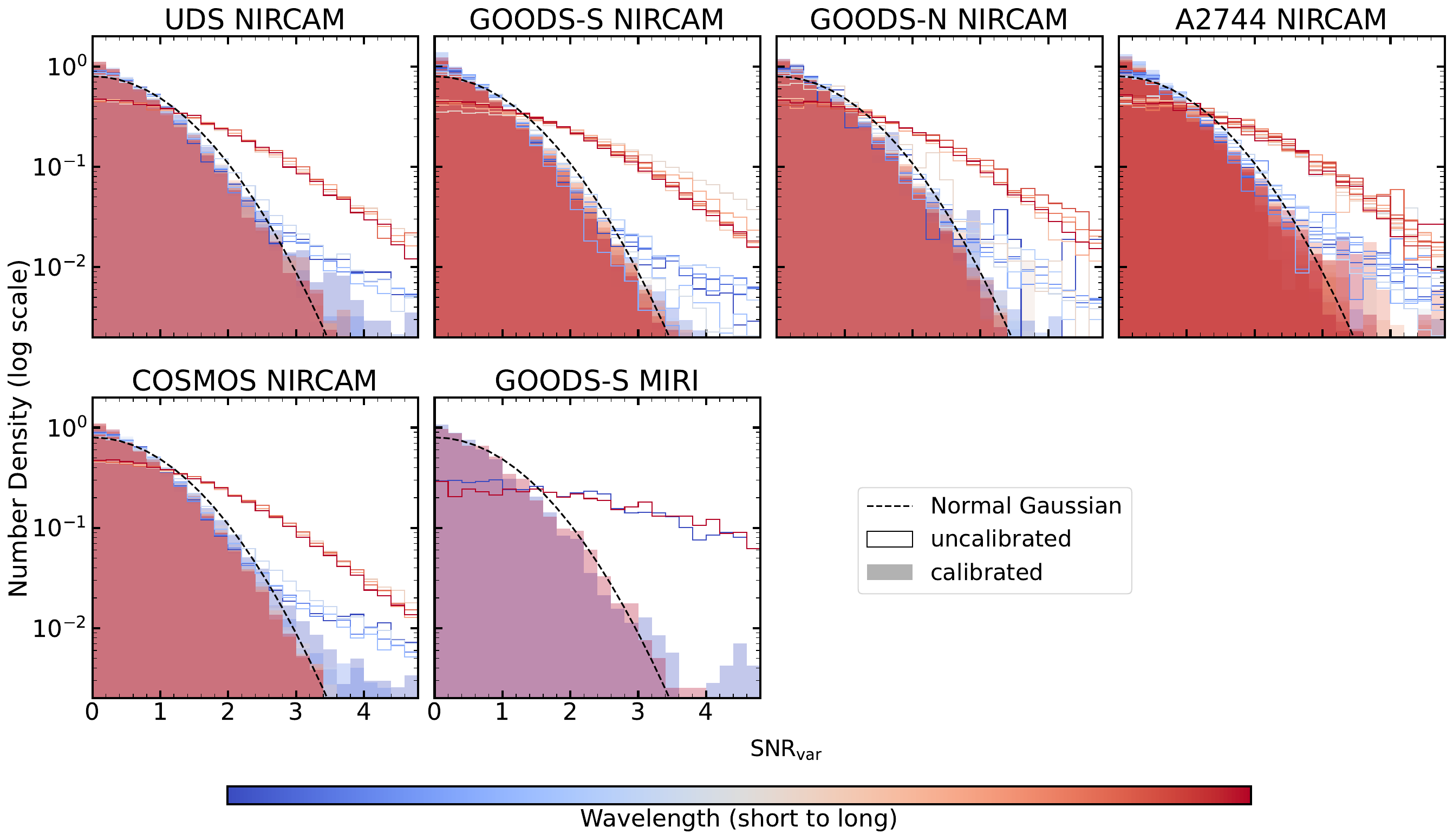}
 \centering
 \caption{Distributions of the $\rm SNR_{\rm var}$ values of all sources for each band in different fields. The shaded histograms represent the calibrated $\rm SNR_{\rm var}$ distributions. Different bands are color-coded as indicated in the color bar at the bottom. The solid line histograms correspond to the $\rm SNR_{\rm var}$ distribution calculated using the uncalibrated $\Delta m$ and errors. The black dashed line shows the standard Gaussian distribution for comparison.
}
 \label{fig:bands_SN_distribution}
\end{figure*}

Using the calibrated $\Delta m$ and errors, we can measure the significance of the variability of an object between 2 \textit{visits} as: 
\begin{equation}
    {\rm SNR_{\rm var}} = \left| \Delta m \right| / \Delta m_{\rm err},
\end{equation}
where $\Delta m$ and $\Delta m_{\rm err}$ are calibrated as described in Section \ref{subsec:correction_of_phot_offset} and Section \ref{subsec:correction_phot_uncertainty}, respectively. For a sample of \hbox{non-variable} sources, the distribution of their ${\rm SNR_{\rm var}}$ should follow a standard Gaussian distribution. We plot the ${\rm SNR_{\rm var}}$ distribution of all sources (with the criteria in Section \ref{subsec:correction_phot_uncertainty}) of each band and each field in Figure \ref{fig:bands_SN_distribution}. One source can contribute multiple ${\rm SNR_{\rm var}}$ values if it was observed in more than two \textit{visits}. In Figure \ref{fig:bands_SN_distribution}, we also plot the ${\rm SNR_{\rm var}}$ distribution calculated using the uncalibrated $\Delta m$ and errors and the standard Gaussian distribution for comparison. The ${\rm SNR_{\rm var}}$ distributions calculated using the calibrated $\Delta m$ and errors are consistent with the standard Gaussian distribution, which confirms the reliability of our error calibration. The uncalibrated ${\rm SNR_{\rm var}}$ distributions deviate from the standard Gaussian distribution, especially for the LW bands, which is consistent with the fact that we need to make a larger error correction to these bands (see Figure \ref{fig:correction_curves}). As we set a minimum error of the measured magnitude as 0.05 mag, the calibrated ${\rm SNR_{\rm var}}$ distributions are more concentrated towards 0 than the standard Gaussian distribution. The calibrated ${\rm SNR_{\rm var}}$ distributions also have tails in their high ${\rm SNR_{\rm var}}$ end. We visually check the light curves and cut-out images of a random sample of the sources contributing to these tails, and find that some of them are real variable sources and others are likely caused by image artifacts like cosmic rays, hot pixels, optical ghosts, etc.

We compare the ${\rm SNR_{\rm var}}$ distribution of the LRD samples to that of fiducial samples to evaluate whether the LRD sample shows significant variability on average. The fiducial sample for each field is drawn from all sources used in Section \ref{subsec:correction_phot_uncertainty} with known AGNs and LRDs removed. We ensure that the fiducial sample has the same magnitude distribution as the LRD sample (all have large KS-test $p$ value) and that the number of sources in the fiducial sample is 20 times the number of LRDs.

\begin{figure*}
\hspace{-0.4cm}
 \includegraphics[width=0.95\textwidth]{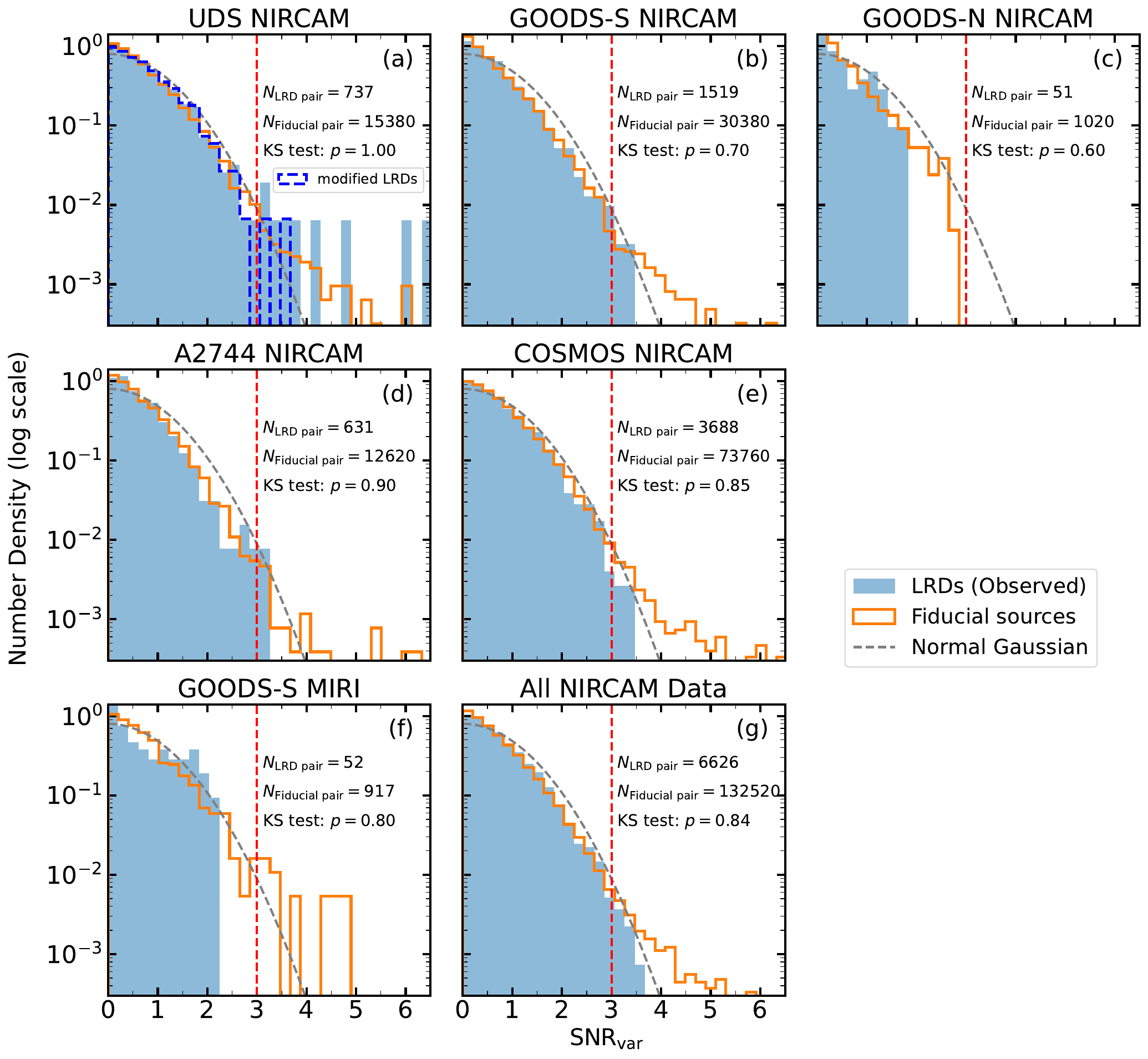}
 \centering
 \caption{$\rm SNR_{\rm var}$ distributions of the LRDs and fiducial sources combing all bands. Panel (a)--(e) shows the distribution for NIRCam data of UDS, GOODS-S, GOODS-N, Abell 2744, and COSMOS fields, respectively. Panel (f) shows the distribution of the MIRI data in GOODS-S. Panel (g) shows the distribution for all NIRCam data. The number of the LRD pairs, number of fiducial pairs, and the KS-test $p$ value are labeled in each panel. The standard Gaussian distribution is represented by a gray dashed line for comparison. The numbers of ${\rm SNR}_{\rm var}$ values used to plot the distributions of the LRD sample and fiducial sample for each field are also labeled in each panel. The red vertical dashed line is $\rm SNR_{\rm var} = 3$. The $\rm SNR_{\rm var}$ distribution of the LRDs is almost the same as that of all other sources, indicating that most LRDs do not show significant variability.}
 \label{fig:all_LRD_SN_distribution}
\end{figure*}

The results of the ${\rm SNR_{\rm var}}$ distribution are shown in Figure \ref{fig:all_LRD_SN_distribution}. The LRD samples all show a nearly consistent ${\rm SNR_{\rm var}}$ distribution with the fiducial samples except for the NIRCam data of the UDS field. The high ${\rm SNR}_{\rm var}$ in the UDS field is mainly caused by 3 LRDs, whose variability is found to be caused by the unique strip-like global background pollution in some \textit{visits}. After removing these three problematic sources, the modified ${\rm SNR}_{\rm var}$ distribution of the LRDs (the dashed blue step distribution in Figure \ref{fig:all_LRD_SN_distribution}a) is also consistent with the fiducial distribution in the UDS field. For the NIRCam data in the GOODS-N field and the MIRI data in the GOODS-S field, there are only 51 and 52 LRD pairs, respectively. Therefore, the expected number of pairs that show ${\rm SNR}_{\rm var}>2$ (corresponding to $<5\%$ for the normalized Gaussian distribution) is smaller than one, which is consistent with our results. The ${\rm SNR}_{\rm var}$ distribution comparison combining the NIRCam data from all fields in Figure \ref{fig:all_LRD_SN_distribution}(g) has the highest statistical significance, and also shows that the LRD distribution and fiducial distribution are almost the same. To conclude, these results indicate that at least most of the LRDs do not show detectable variability. A similar non-detection was also reported by \citet{2025ApJ...983L..26T}, who used a combination of HST and JWST observations to study the variability of 21 LRDs over observed-frame time intervals ranging from 6 to 11 years.

\subsection{No Significant Variability of the LRDs with Broad Emission Lines}
\label{subsec:no_vari_BL_LRD}

\begin{figure}[ht]
\hspace{-0.4cm}
 \includegraphics[width=0.47\textwidth]{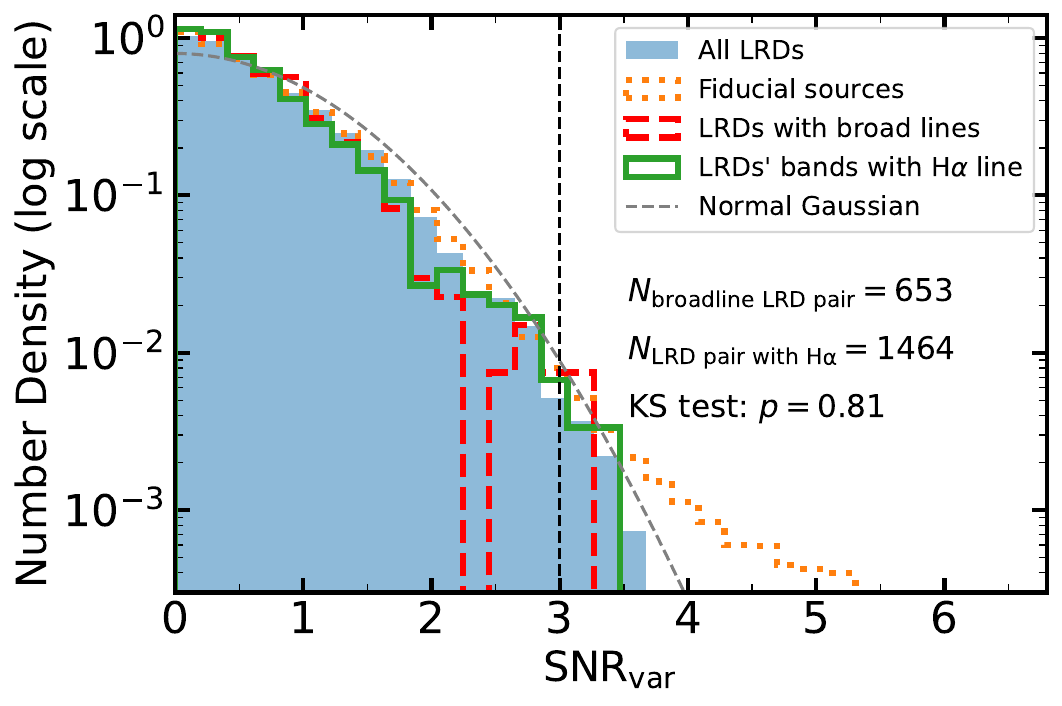}
 \centering
 \caption{$\rm SNR_{\rm var}$ distribution of the LRDs that show broad lines (the red dashed line). For comparison, the green solid line represents the $\rm SNR_{\rm var}$ distribution of all LRDs that likely have H$\alpha$ line emission suggested by their spec-$z$/photo-$z$ measurements. The $\rm SNR_{\rm var}$ distribution (blue) of the whole LRD sample is the same as shown in Figure \ref{fig:all_LRD_SN_distribution}(g). The fiducial $\rm SNR_{\rm var}$ distribution is drawn to have the same magnitude distribution as the broad-line LRD sample, with the KS-test $p$ value of these two sample labeled.} 
 \label{fig:LRD_BL_variability}
\end{figure}

Among the 314 LRDs that have multi-epoch observation, 27 of them show broad Balmer lines in the literature \citep{2023ApJ...959...39H,2024arXiv240610341A,2024ApJ...964...39G,2024arXiv240403576K,2024ApJ...975..178K,2024arXiv240500504M,2024ApJ...963..129M,2024arXiv240302304W,2024ApJ...968...34W}. These sources are considered to be the most reliable AGN candidates within the LRD sample. We plot the ${\rm SNR}_{\rm var}$ distribution of these broad-line LRDs in Figure \ref{fig:LRD_BL_variability}. Their ${\rm SNR}_{\rm var}$ distribution also shows no sign of variability compared to the fiducial distribution and the standard Gaussian distribution. 

The broad H$\alpha$ emission lines from some LRDs have high equivalent widths \citep[EWs; e.g.,][]{2024A&A...691A.145M,2024arXiv240302304W,2024ApJ...969L..13W}, which is similar to or higher than that expected for AGNs  \citep[e.g., the $L_{\rm H\alpha}$--$L_{\rm 5100\AA}$ relation in][]{2005ApJ...630..122G}. This suggests that the continua of these LRDs should be dominated by AGNs at least in the bands that contain broad H$\alpha$ emission, because otherwise their EWs would be lower. Therefore, we also plot the ${\rm SNR}_{\rm var}$ distribution of the filters that are expected to contain H$\alpha$ line based on their spec-$z$/photo-$z$ for all LRDs in Figure \ref{fig:LRD_BL_variability}. This distribution also indicates no significant variability.

\subsection{Eight Significantly Variable LRD Candidates}
\label{subsec:individual_LRD_variability}

Although most of the LRDs do not show significant variability, there are still 10 LRDs that show variability with ${\rm SNR}_{\rm var}>3$ in at least one \textit{visit} pair. Our visual inspection reveals that the variability of two of them is likely caused by the contamination of a nearby bright source and a cosmic ray, respectively. After removing these two sources, we identify eight variable LRD candidates. Table \ref{tab:vari_LRDs} summarizes the basic information of these sources. Their maximum variability amplitudes are 0.24 -- 0.82 mag. A2744-14 (GLASS 150029 in \citealt{2023ApJ...959...39H}) was also analyzed by \citet{2024arXiv240704777K}, but they reported no significant variability. We find that A2744-14 shows variability of ${\rm SNR}_{\rm var}>3$ in the F360M bands. This band is not included in \citet{2024arXiv240704777K}. 

We can usually neglect the impact of the PSF spatial and temporal variations on the above identification procedure, as we will demonstrate in Section \ref{subsec:effect_of_spatial_variation}. We further inspect the PSF variations for each epoch pair where LRD variability was detected earlier. The details are described in Appendix \ref{appdx:psf_vari}. Our analysis confirms that the effects of PSF variations are negligible for the eight sources.

The color distribution of the eight variable LRD candidates is shown in Figure \ref{fig:color_color}. Compared with the whole LRD sample, their colors tend to be relatively bluer. This suggests that the red color of those reddest LRDs may be caused by stellar emission, while LRDs that are relatively blue may have less host contamination and thus exhibit the intrinsic variability of AGNs.

Considering that the ${\rm SNR}_{\rm var}$ distribution of all LRDs in Figure \ref{fig:all_LRD_SN_distribution}(g) is in a good agreement with the standard Gaussian distribution, it is likely that some of these variable LRD candidates are false detections. However, we suggest that at least two of them (COS-584 and COS-593) are reliable detections because of their relatively large variability amplitudes and coordinated trend of variability in different bands. These two sources both come from the PRIMER-COSMOS field that has relatively more observations per source (there are 3688/6631 LRD pairs for 114/314 LRDs in this field). In addition, different bands in this field often have nearly simultaneous observations; other fields usually lack such observations. In the following subsections, we will provide detailed information about these two LRDs, including their multi-band light curves, multi-band and multi-epoch cut-out images, surface brightness (SB) profiles, and SEDs.

\subsubsection{COS-584}
\label{subsubsec:LID584}

\begin{figure*}[]
\centering
\begin{minipage}{0.9\columnwidth}
      \hspace{-0.6cm}
      \includegraphics[clip,trim=0 0.cm 0 0cm, width=1.02\columnwidth]{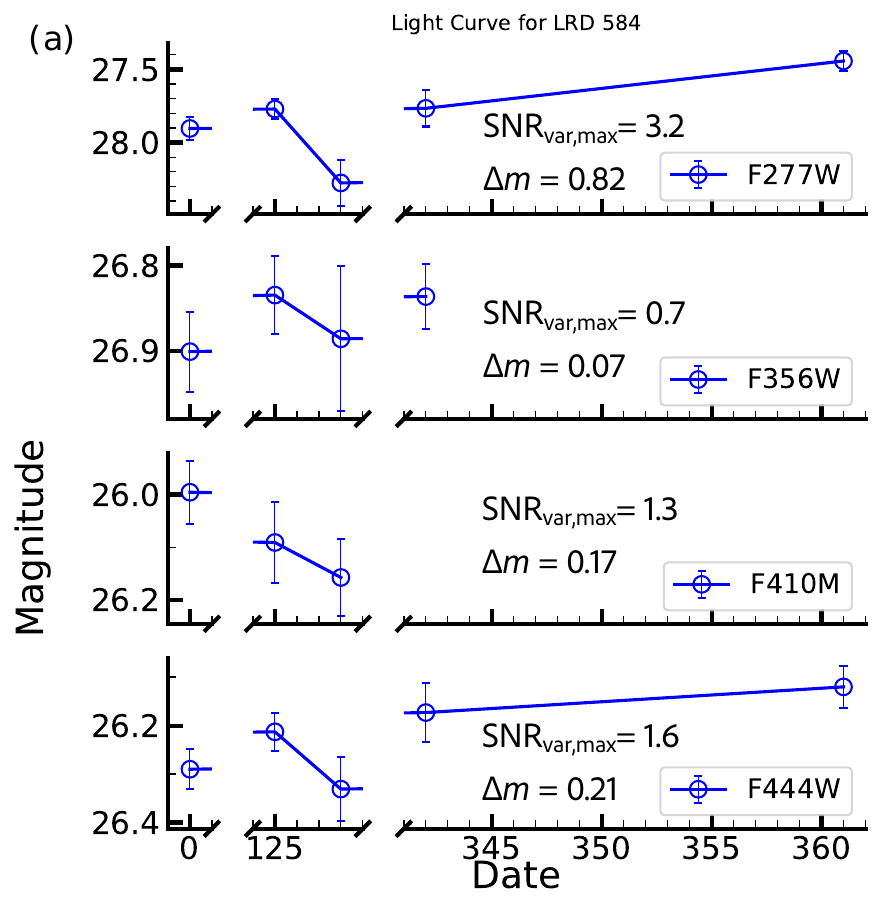}

      \includegraphics[clip,trim=0 0.cm 0 0cm, width=1.02\columnwidth]{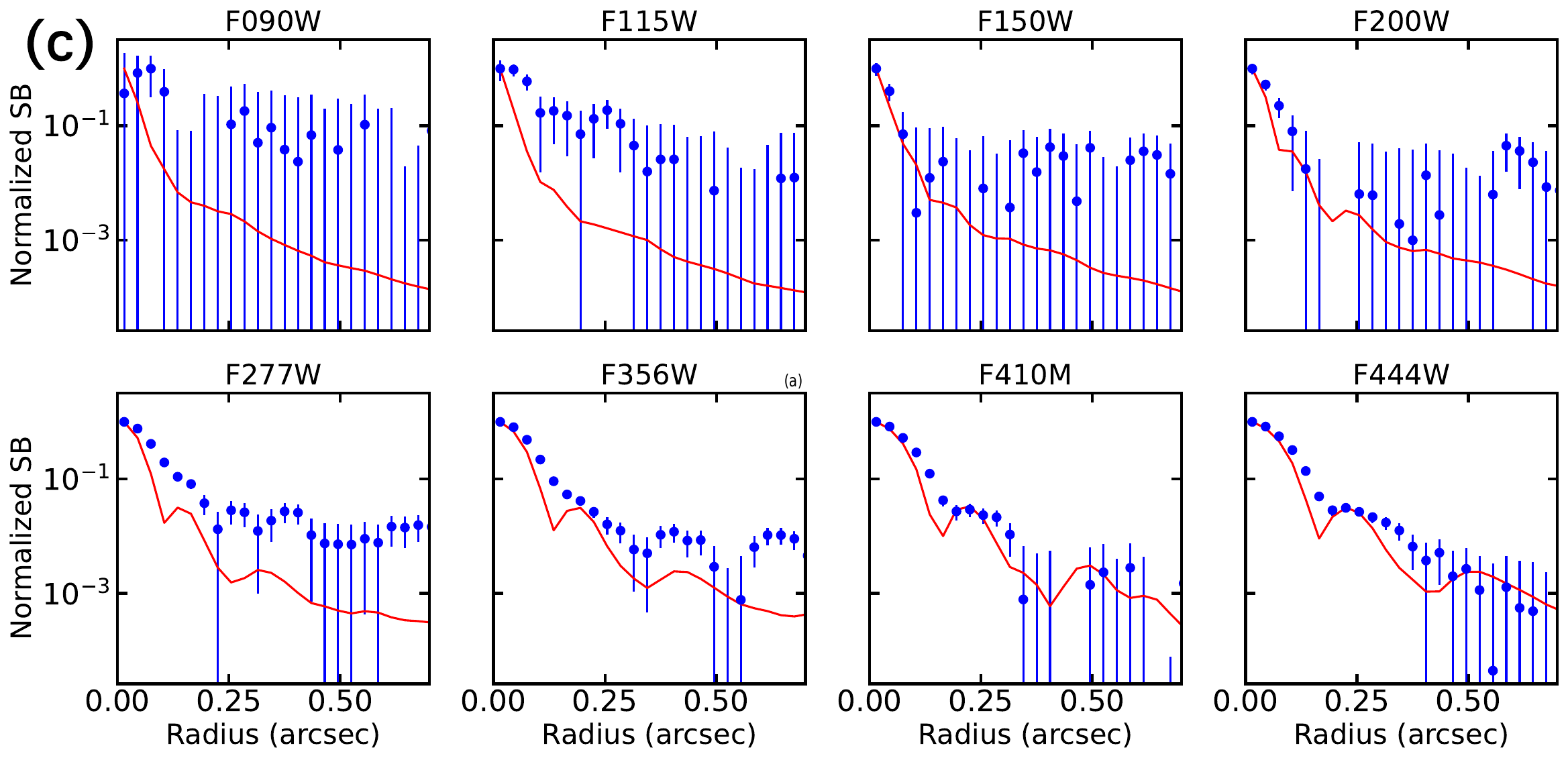}
\end{minipage}
\begin{minipage}{1.0\columnwidth}
      \hspace{0.5cm}
      \vspace{-0.3cm}
      \includegraphics[clip,trim=0 0.cm 0.5cm 0.cm, width=0.9\columnwidth]{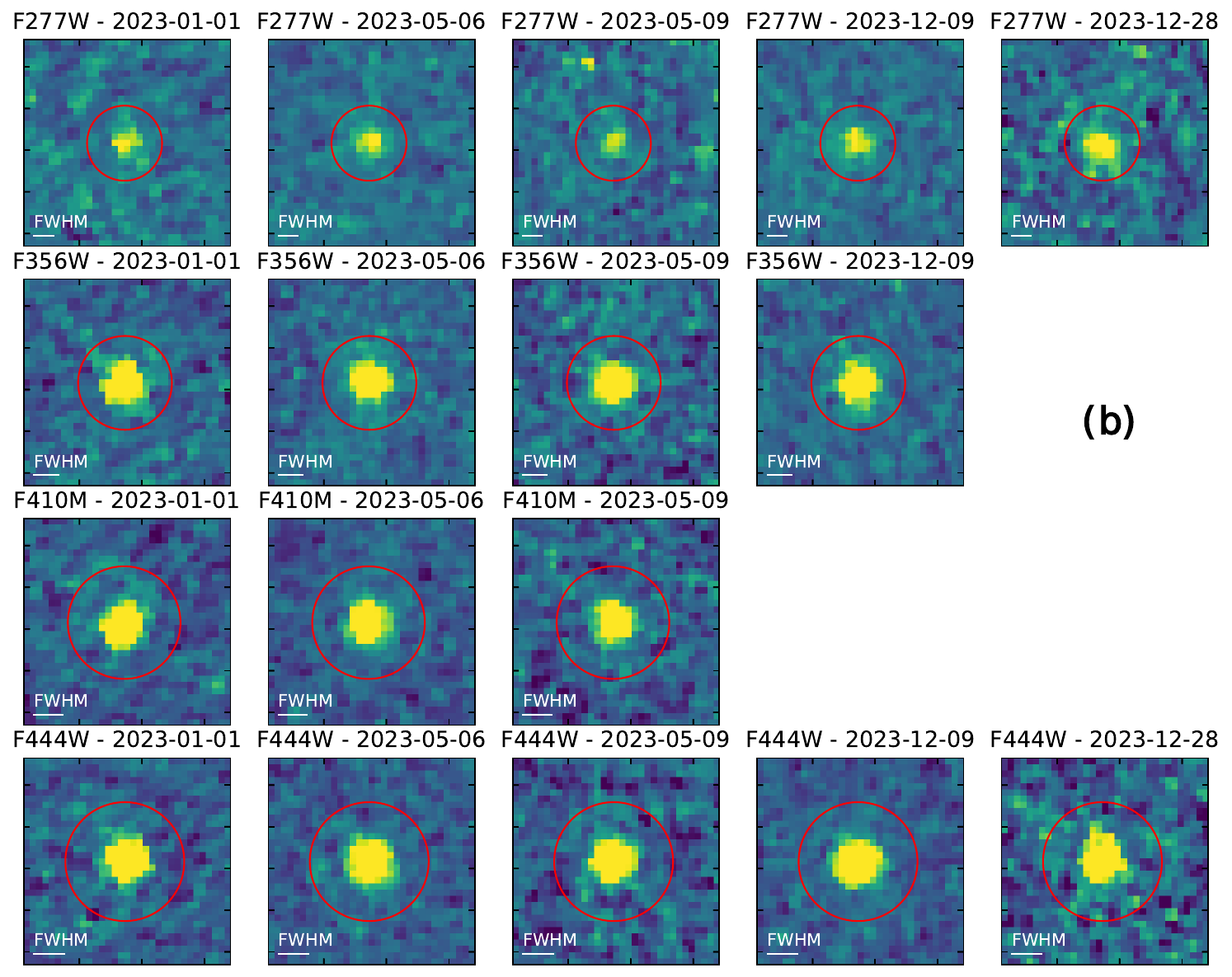}

      \includegraphics[clip,trim=0 0.cm 0 0.cm, width=1.0\columnwidth]{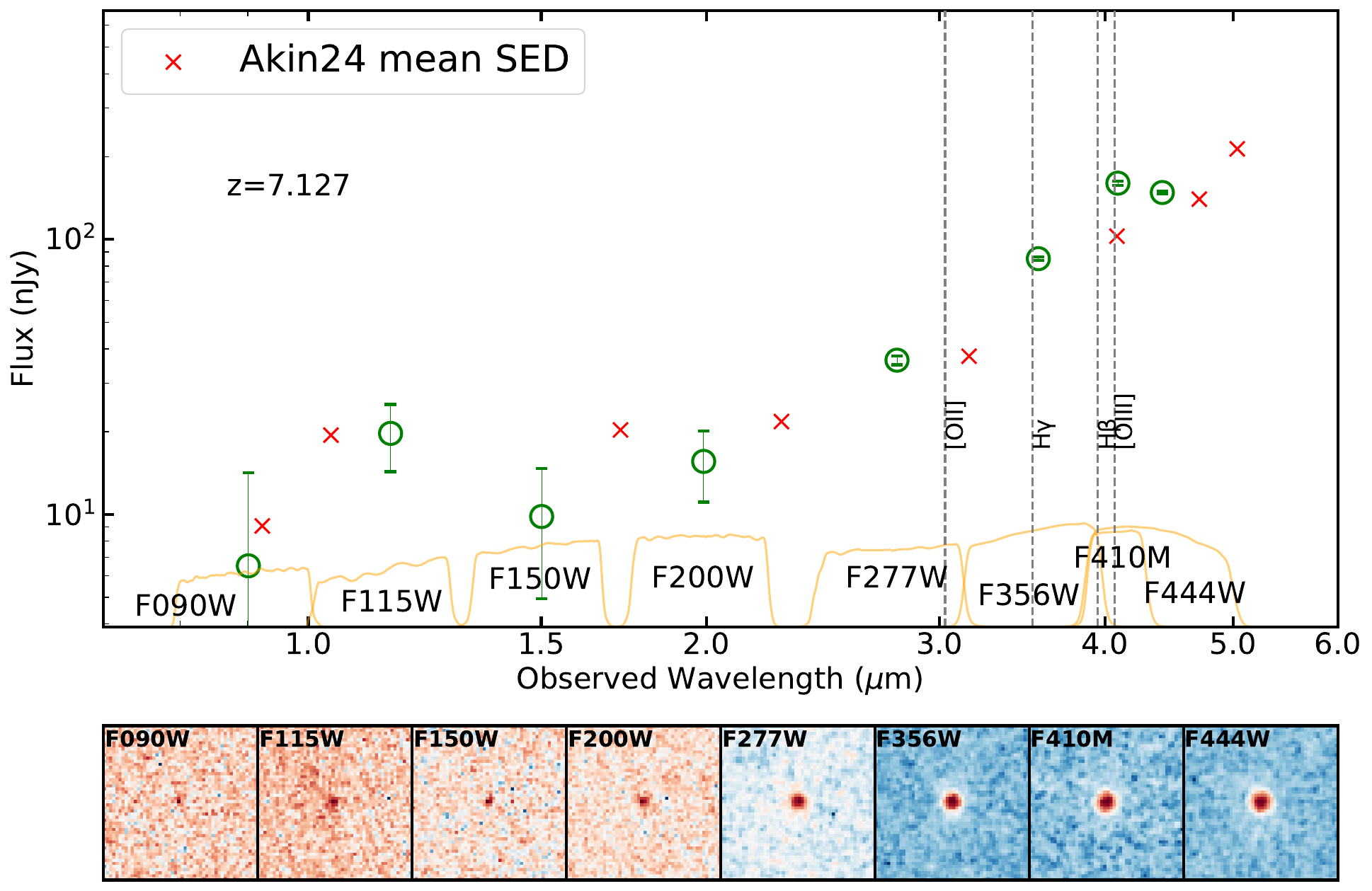}
\end{minipage}
\caption{A kaleidoscope of COS-584. (a) Multi-band light curves. The $x$-axis shows the observer-frame dates, starting from the date of the first observation. The maximal ${\rm SNR}_{\rm var}$ and the corresponding $\Delta m$ are noted for each band. (b) Multi-epoch cut-out images with a size of $1''$. The band and observation date of each image are noted on the top. The red circles represent the photometric aperture in Section \ref{subsec:photometry}. Images of the same band use the same scale. (c) SB profiles at different NIRCam bands (blue error bars). The SB profiles are normalized at the peak. The red curves show the SB profiles of the JWST PSFs obtained from \textit{WebbPSF}. (d) SED and mosaic cut-out images in different NIRCam bands. The average SED of LRDs from \citet{2024arXiv240610341A} are shown for comparison. Main emission lines are indicated by the vertical dotted lines.}
\label{fig:lc_img_584}
\end{figure*}

COS-584 was selected by \citet{2024arXiv240403576K} and \citet{2024ApJ...968...38K} as a LRD at a photo-$z$ of 7.127. It brightens by 0.82 mag in the F277W band between 2023-05-09 and 2023-12-28, which is the largest variability amplitude of the eight variable LRDs. Its light curves in other bands show a coordinated trend with F277W but with smaller variability amplitudes and ${\rm SNR}_{\rm var}$, as shown in Figure \ref{fig:lc_img_584}(a). This multi-band coordination confirms the variability of COS-584. It shows a smaller variability amplitude in longer-wavelength bands, which is consistent with the expectation of AGNs. 

\begin{figure*}
\centering
\begin{minipage}{0.99\columnwidth}
    
  \includegraphics[width=0.99\textwidth]{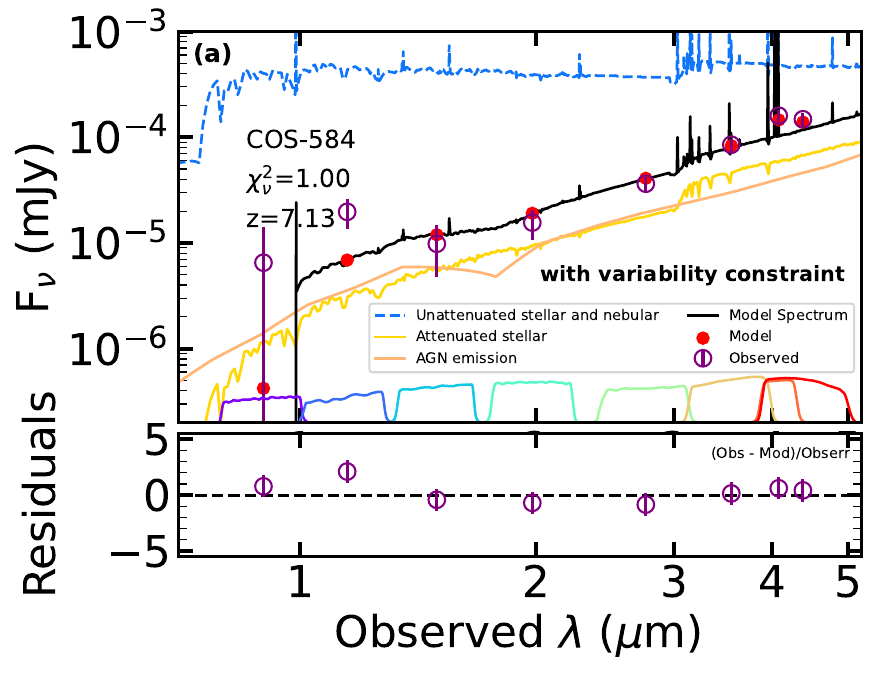}

\end{minipage}
\begin{minipage}{0.99\columnwidth}

  \includegraphics[width=0.99\textwidth]{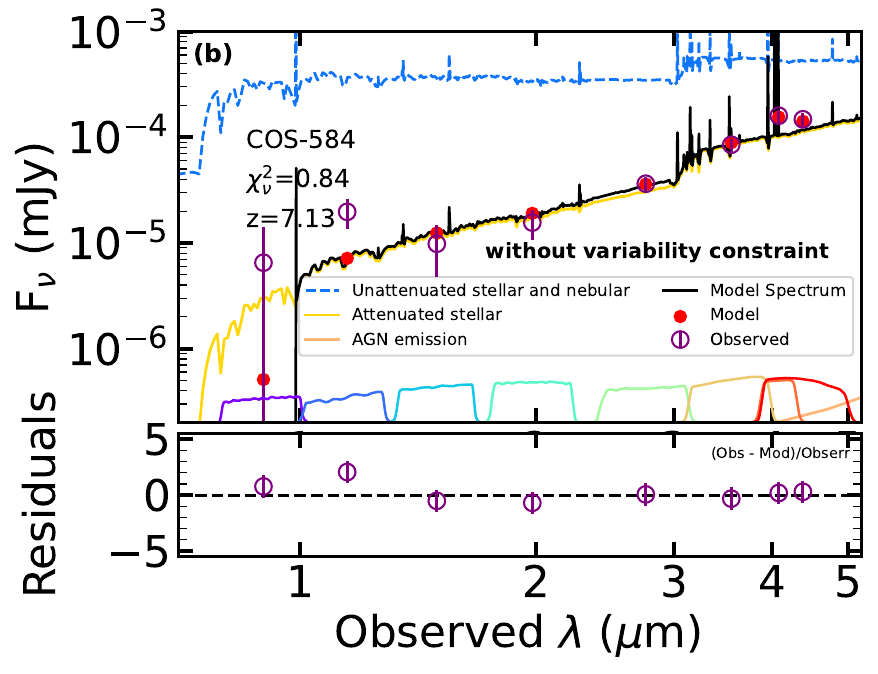}
\end{minipage}

\caption{Best-fit SED models of COS-584 (a) with and (b) without the constraint of the AGN fraction set by the maximal variability amplitude. The bottom panel displays the relative residuals between the observed and inferred photometry.}
\label{fig:sedfitting_584}
\end{figure*}

We model the multi-band SED of COS-584 with SED fitting. We assume that there are two components in this system, including a central AGN that causes variability and its host galaxy. We try to break the degeneracy between the two components using the variability amplitude as a constraint. The variability amplitude sets a lower limit on the AGN fraction as $\frac{f_{\rm mosaic} - f_{\rm min}}{f_{\rm min}} = 0.36$ in the F277W-band mosaic image, assuming $f_{\rm min}$ is solely from the galaxy component. We also use \texttt{GalfitM} \citep{2013MNRAS.430..330H,2013MNRAS.435..623V} to perform multi-band image fitting and decompose the system into a central PSF component and a \sersic\ galaxy component (see Appendix \ref{appxsec:galfitm} for details). The fraction of the PSF component is treated as an upper limit on the AGN fraction, as part of the stellar component may be compact and unresolved at high redshift \citep[e.g.,][]{2023Natur.619..716C,2023ApJ...951...72O}. With these constraints, we fit the JWST/NIRCam SED of COS-584 with a modified version of \texttt{CIGALE} \citep[v2025, ][]{2005MNRAS.360.1413B,2019A&A...622A.103B}. The SED fitting configuration is detailed in Appendix \ref{appxsec:sedfitting}. For comparison, we also perform SED fitting without the variability constraint {but with the morphological constraints}. The best-fit results are shown in Figure \ref{fig:sedfitting_584}. With all constraints applied (Figure \ref{fig:sedfitting_584}a), the best-fit SED shows similar stellar/galaxy and AGN contributions across the observed wavelength range. The host galaxy component increases toward longer wavelengths, which can dilute the intrinsic AGN variability and partly explain the reduced variability amplitude at longer wavelengths. Without the variability constraint (Figure \ref{fig:sedfitting_584}b), the best-fit SED is entirely dominated by the galaxy component, highlighting the importance of the variability information for breaking the degeneracy in the SED fitting.

\begin{longrotatetable}
\begin{deluxetable*}{CCCCCCCCCCCCCCCC}
\centering
\tablecaption{LRDs that show variability of ${\rm SNR}_{\rm var}>3$}\label{tab:vari_LRDs}
\tablewidth{670pt}
\setlength\tabcolsep{3pt}
\colnumbers
\tablehead{
  \colhead{LID} & 
  \colhead{R.A.} & 
  \colhead{Decl.} & 
  \colhead{$M_{\rm UV}$} & 
  \colhead{$z$} & 
  \colhead{ztype} & 
  \colhead{isBL} & 
  \colhead{$\log M_{\rm BH}$} & 
  \colhead{${\rm SF}_{\infty}$} & 
  \colhead{${\rm SNR}_{\rm var}$} & 
  \colhead{$\Delta m$} & 
  \colhead{$\rm N_{band}$} & 
  \colhead{$\rm N_{obs}$} & 
  \colhead{band} & 
  \colhead{date} & 
  \colhead{ref} 
}

\decimals
\startdata
\textnormal{A2744-}14&3.577167&-30.422583&-19.2&4.583&1&1&6.6&<-0.88&3.2&0.28&12&34&\textnormal{F360M}&\textnormal{2023-10-31--2023-11-12}&\textnormal{H23}\\
\hline
\textnormal{UDS-}108&34.363515&-5.176964&-19.2&4.052&1&0&...&<-1.24&3.6&0.25&6&19&\textnormal{F200W}&\textnormal{2023-08-07--2023-08-09}&\textnormal{K24a/K24b}\\
\hline
\textnormal{UDS-}147&34.411921&-5.144804&-18.2&7.782&1&0&...&<-0.26&3.2&0.71&2&4&\textnormal{F444W}&\textnormal{2023-08-07--2023-08-09}&\textnormal{K24a}\\
\hline
\textnormal{GS-}258&53.159333&-27.811749&-17.6&4.942&0&0&...&<-0.98&3.4&0.47&7&24&\textnormal{F410M}&\textnormal{2022-10-03--2022-10-04}&\textnormal{K24a/K24b}\\
\hline
\textnormal{COS-}538&150.111076&2.245270&-19.3&7.135&0&0&...&<-1.42&3.4&0.24&6&36&\textnormal{F410M}&\textnormal{2023-04-27--2023-05-22}&\textnormal{K24a/K24b/A24}\\
\hline
\textnormal{COS-}584&150.149360&2.252564&-17.7&7.127&0&0&...&<-0.14&3.2&0.82&4&17&\textnormal{F277W}&\textnormal{2023-05-09--2023-12-28}&\textnormal{K24a/K24b}\\
\hline
\textnormal{COS-}593&150.152769&2.273189&-17.5&5.227&0&0&...&<-0.02&3.3&0.61&4&21&\textnormal{F356W}&\textnormal{2023-01-03--2023-12-09}&\textnormal{K24a/K24b}\\
\hline
\textnormal{COS-}621&150.170181&2.388390&-20.1&5.367&0&0&...&<-1.68&3.1&0.47&8&35&\textnormal{F150W}&\textnormal{2023-01-03--2023-12-29}&\textnormal{K24a/K24b}\\
\enddata
\tablecomments{
Col. (1): LID assigned in Section \ref{sec:sample} according to the R.A. order.
Cols. (2) and (3): Right Ascension and Declination.
Col. (4): UV absolute magnitude adopted from \citet{2024arXiv240403576K} and \citet{2023ApJ...959...39H}.
Col. (5): Redshift adopted from \citet{2024arXiv240403576K} and \citet{2023ApJ...959...39H}.
Col. (6): Type of Redshift: 1 denotes spec-$z$ and 0 denotes photo-$z$.
Col. (7): Label indicates the presence of broad emission line: 1 means the LRD is revealed to have broad emission line and 0 means the LRD do not have spectroscopic observation yet.
Col. (8): Black hole mass of A2744-14 adopted from \citet{2024arXiv240403576K}.
Col. (9): Upper limit of ${\rm SF}_{\infty}(4000\AA)$ derived as described in Section \ref{subsec:constraint_amplitude}.
Col. (10): Maximal ${\rm SNR}_{\rm var}$ value calculated as described in Section \ref{subsec:variability_of_LRD}.
Col. (11): The magnitude difference corresponds to the maximal ${\rm SNR}_{\rm var}$ value.
Col. (12): Total number of bands that have multi-epoch detections.
Col. (13): Total number of multi-epoch detections with all bands combined.
Col. (14): The band in which the maximal ${\rm SNR}_{\rm var}$ value measured.
Col. (15): The date pair on which the maximum SN value measured.
Col. (16): The reference that select the object as a LRD or H$\alpha$ emitter. H23: \citet{2023ApJ...959...39H}; A24: \citet{2024arXiv240610341A}; K24a: \citet{2024arXiv240403576K}; K24b: \citet{2024ApJ...968...38K}.
}
\end{deluxetable*}
\end{longrotatetable}

\begin{figure*}[ht]
\centering
\begin{minipage}{0.9\columnwidth}

      \hspace{-0.6cm}
      \includegraphics[clip,trim=0 0.cm 0 0cm, width=1.02\columnwidth]{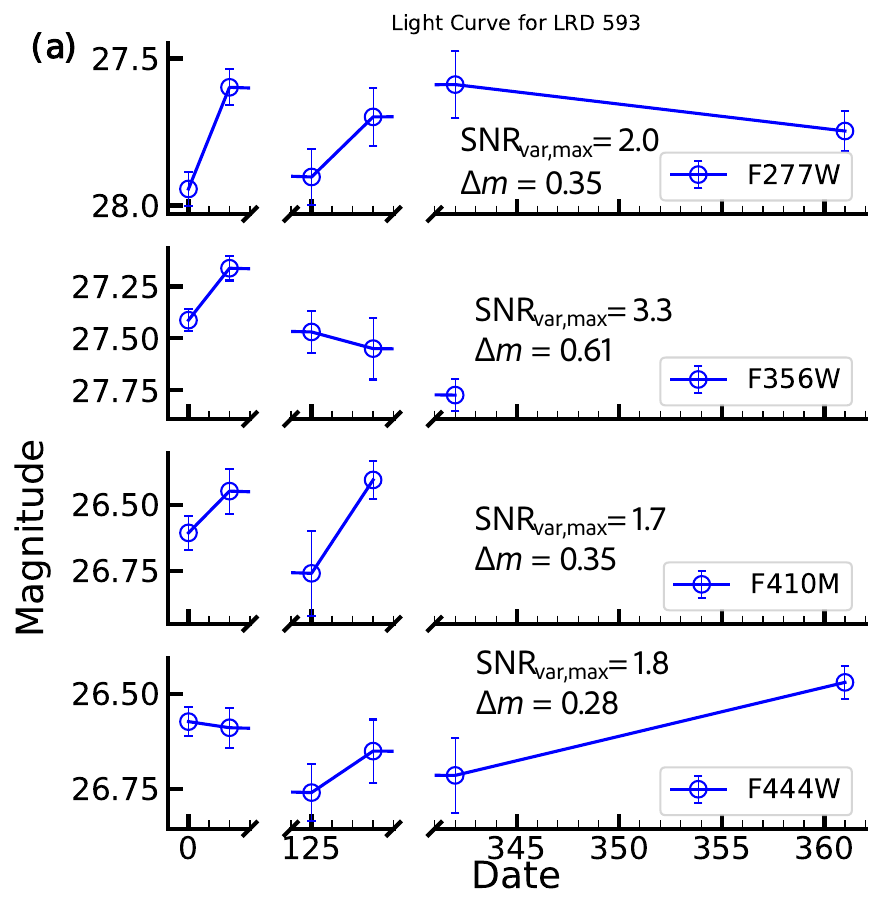}

      \includegraphics[clip,trim=0 0.cm 0 0cm, width=1.02\columnwidth]{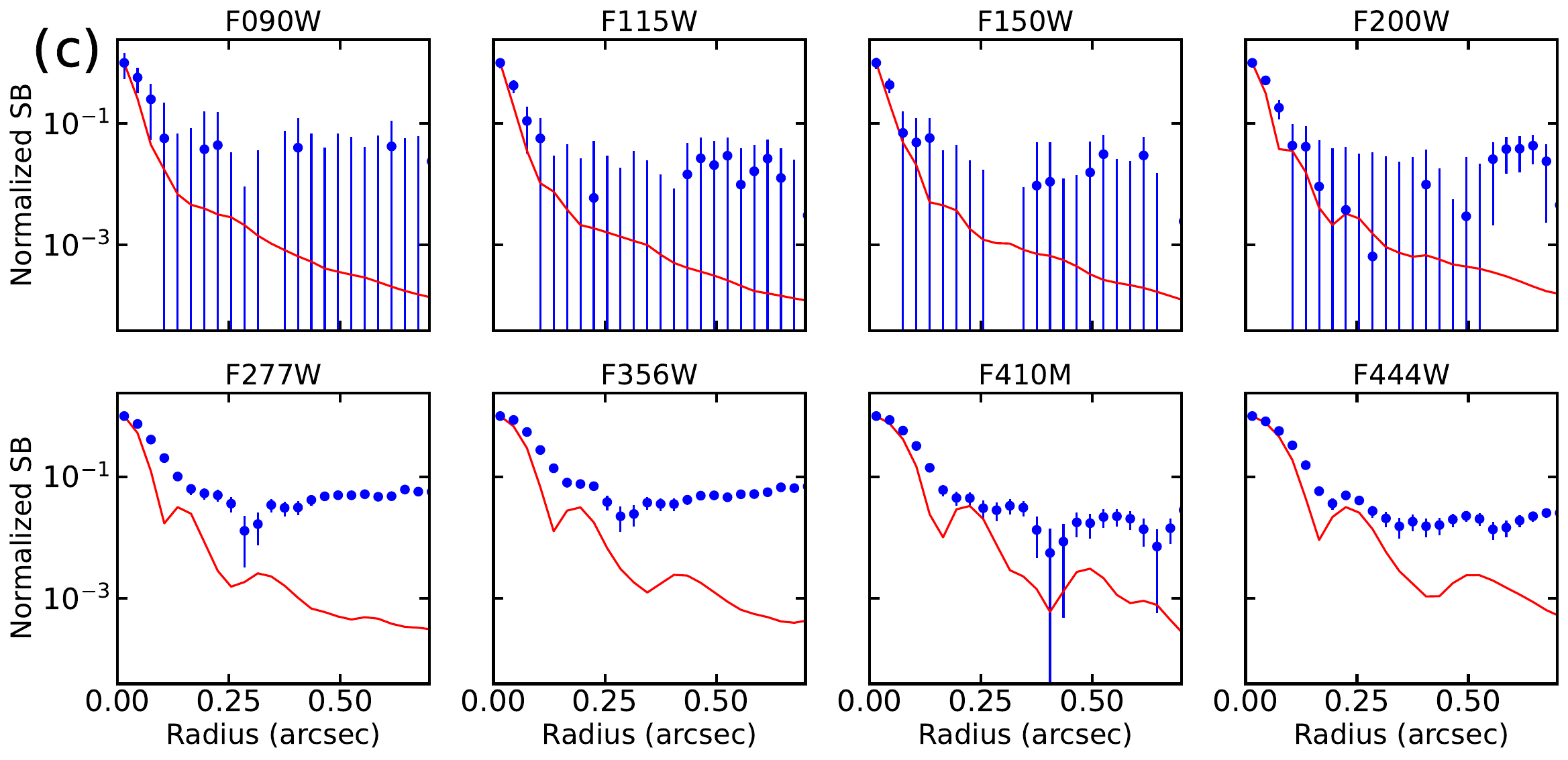}

\end{minipage}
\begin{minipage}{1.0\columnwidth}
      \hspace{0.5cm}
      \vspace{-0.3cm}
      \includegraphics[clip,trim=0 0.cm 0.5cm 0.cm, width=0.9\columnwidth]{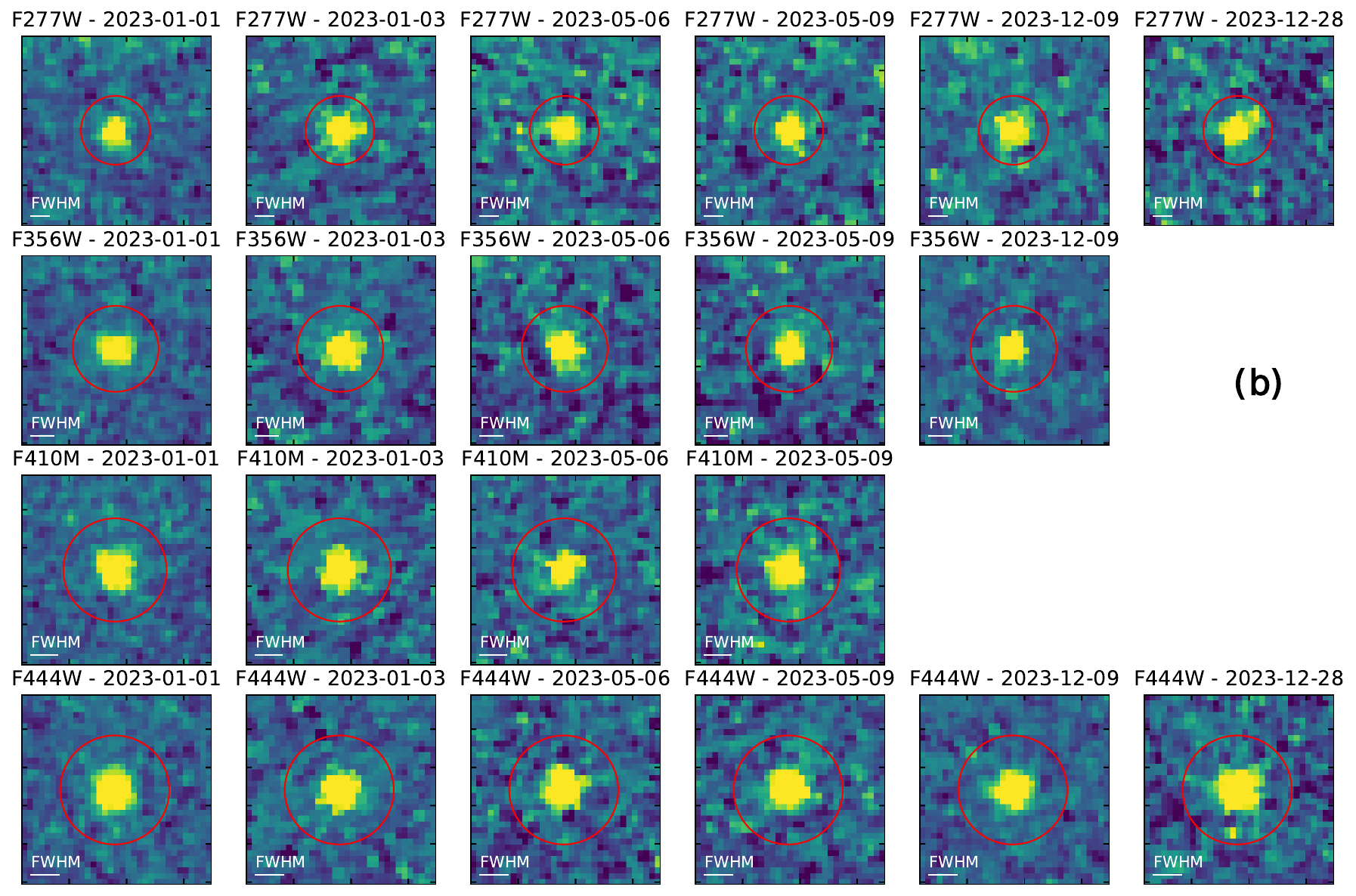}

      \includegraphics[clip,trim=0 0.cm 0 0.cm, width=1.0\columnwidth]{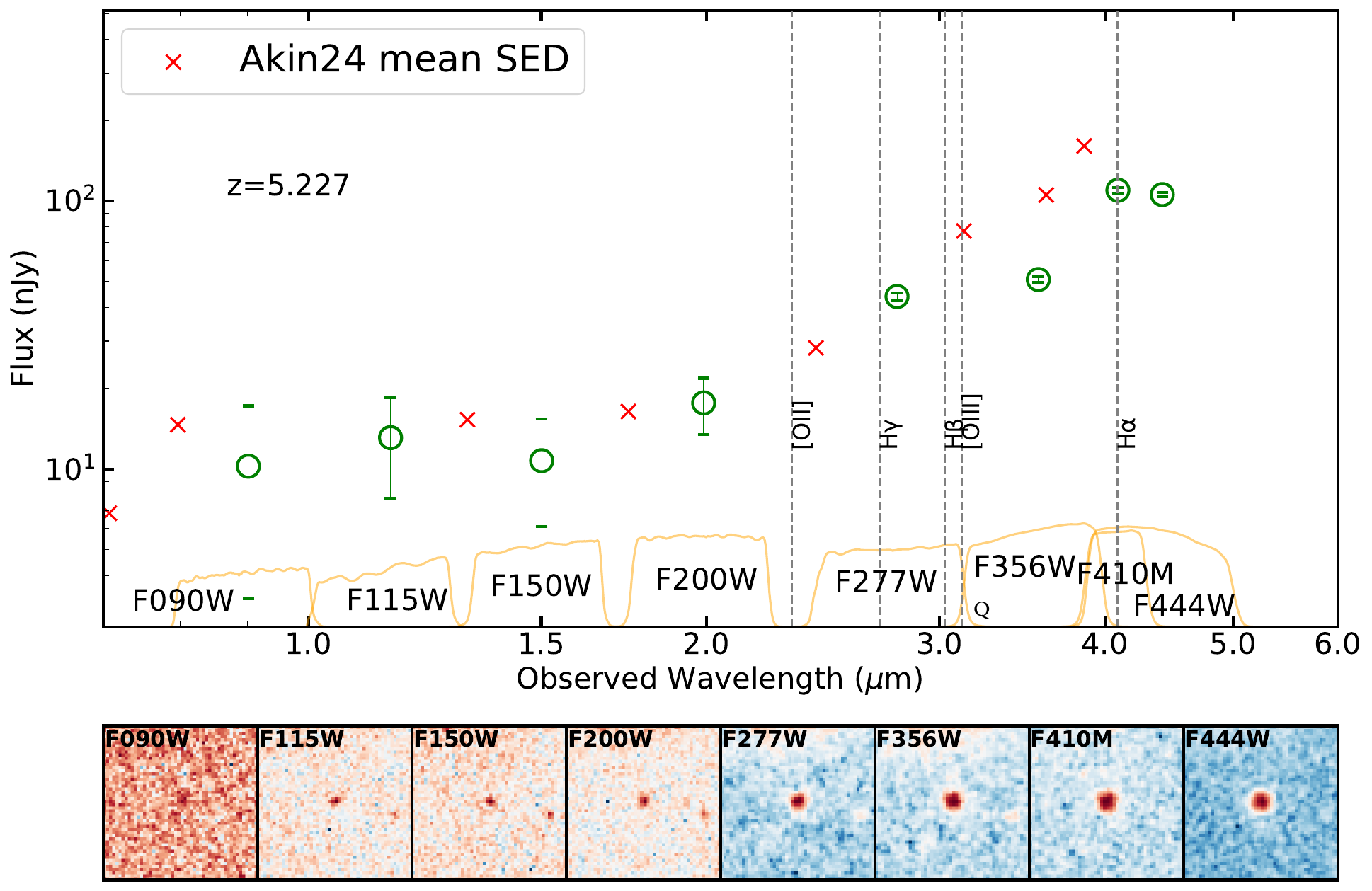}
\end{minipage}
\caption{Same as Figure \ref{fig:lc_img_584}, but for COS-593.}
\label{fig:lc_img_593}
\end{figure*}

\begin{figure}[ht]
\hspace{-0.4cm}
 \includegraphics[width=0.49\textwidth]{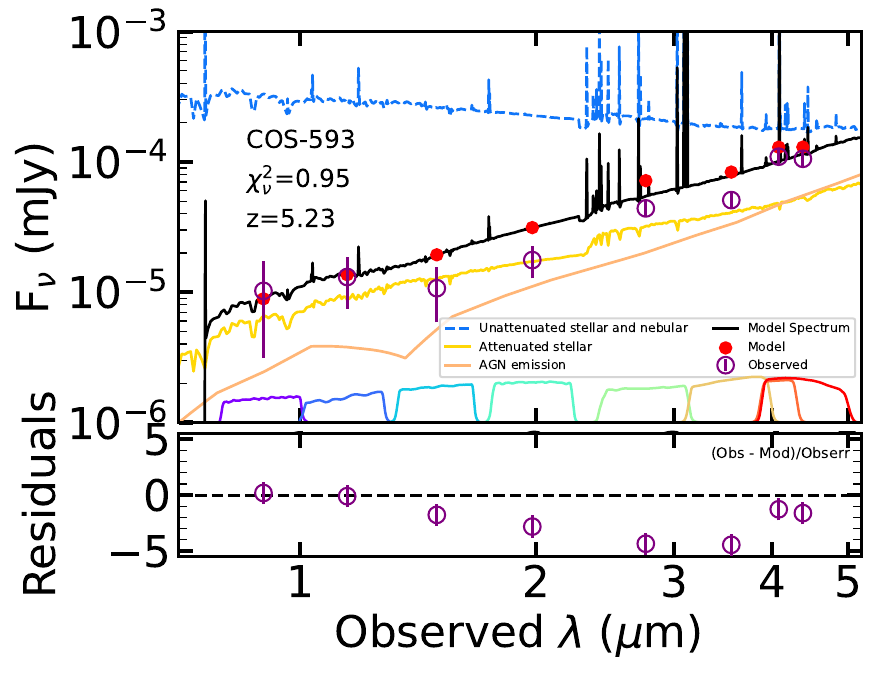}
 \centering
 \caption{Same as Figure \ref{fig:sedfitting_584}, but for COS-593. The fits with and without the constraint from the variability information give the same result.}
 \label{fig:SEDfitting_593}
\end{figure}

\subsubsection{COS-593}
\label{subsubsec:LID593}

COS-593 was also selected by \citet{2024arXiv240403576K}  and \citet{2024ApJ...968...38K} as a LRD (their ID 22990 and ID 17813, respectively). Its photo-$z$ is 5.2 in both works. COS-593 shows a maximum variability of $\Delta m = 0.61$ mag in the F356W band with ${\rm SNR}_{\rm var}=3.3$ between the observations on 2023-01-03 and 2023-12-09.

Although the maximum ${\rm SNR}_{\rm var}$ values in the other three bands do not exceed 3, they all show a consistent variability trend. However, the variability trend of F356W is different from those in the other three bands. As shown in Figure \ref{fig:lc_img_593}(d), the F277W, F410M, and F444W filters are all expected to enclose the H$\alpha$ or H$\beta$ emission lines for a redshift of 5.2, while F356W is line-free. The different trends of the light curves are likely caused by the time lag between the emission line and continuum variability, which is expected for AGNs. The continuum luminosity $L_{5100}$ of the AGN from the best-fit SED model (see below) is $5\times 10^{42} \rm ~erg~s^{-1}$. Using the size-luminosity relation in \citet{2016ApJ...825..126D}, the time lag of the H$\beta$ line is estimated to be $\sim 10$ days in the rest frame, corresponding to $\sim 60$ days in the observed frame. Assuming that the time scale of H$\alpha$ lag is similar to H$\beta$, the different trends of the light curves are consistent with the scenario of the time lag, although the sampling of the light curves is sparse.

The SED of COS-593 is bluer than the average LRD SED from \citet{2024arXiv240610341A}. We perform the same SED modeling as we did for COS-584. The constraint from the variability information sets a lower limit of $>24\%$ on the AGN fraction in the F356W band. The morphological constraints are also applied using the same method. For COS-593, the SED fitting procedure with and without the constraints {from the variability} yields the same result shown in Figure \ref{fig:SEDfitting_593}. This is because the AGN fraction constraint is not stringent enough to affect the result. {The morphological constraints reduce the model’s flexibility, leading to best-fit model with systematically lower fluxes compared to the observed photometry.} The best-fit SED is dominated by the galaxy component in the rest-frame UV, and the AGN fraction increases with wavelength. This is consistent with the decrease of the extended emission towards longer wavelengths as demonstrated by the SB profiles.

\section{Discussion}
\label{sec:discussion}

\subsection{Effect of the PSF Spatial Variation}
\label{subsec:effect_of_spatial_variation}

It is not realistic to achieve a perfect PSF that remains constant over time and across different positions on the telescope focal plane. Recent studies have demonstrated that the JWST NIRCam PSF exhibits significant spatial variations of up to approximately 15--20\% \citep[e.g.,][]{2022MNRAS.517..484N,2024ApJ...962..139Z}, as well as temporal variations around 3--4\% \citep[e.g.,][]{2022MNRAS.517..484N}. The temporal variations are small and thus nearly negligible given that we have corrected the relative photometric zero point offsets across different \textit{visits} (see Section \ref{subsec:correction_of_phot_offset}). The spatial variations are relatively large and may affect our results.

\begin{figure*}[ht]
\hspace{-0.4cm}
 \includegraphics[width=0.9\textwidth]{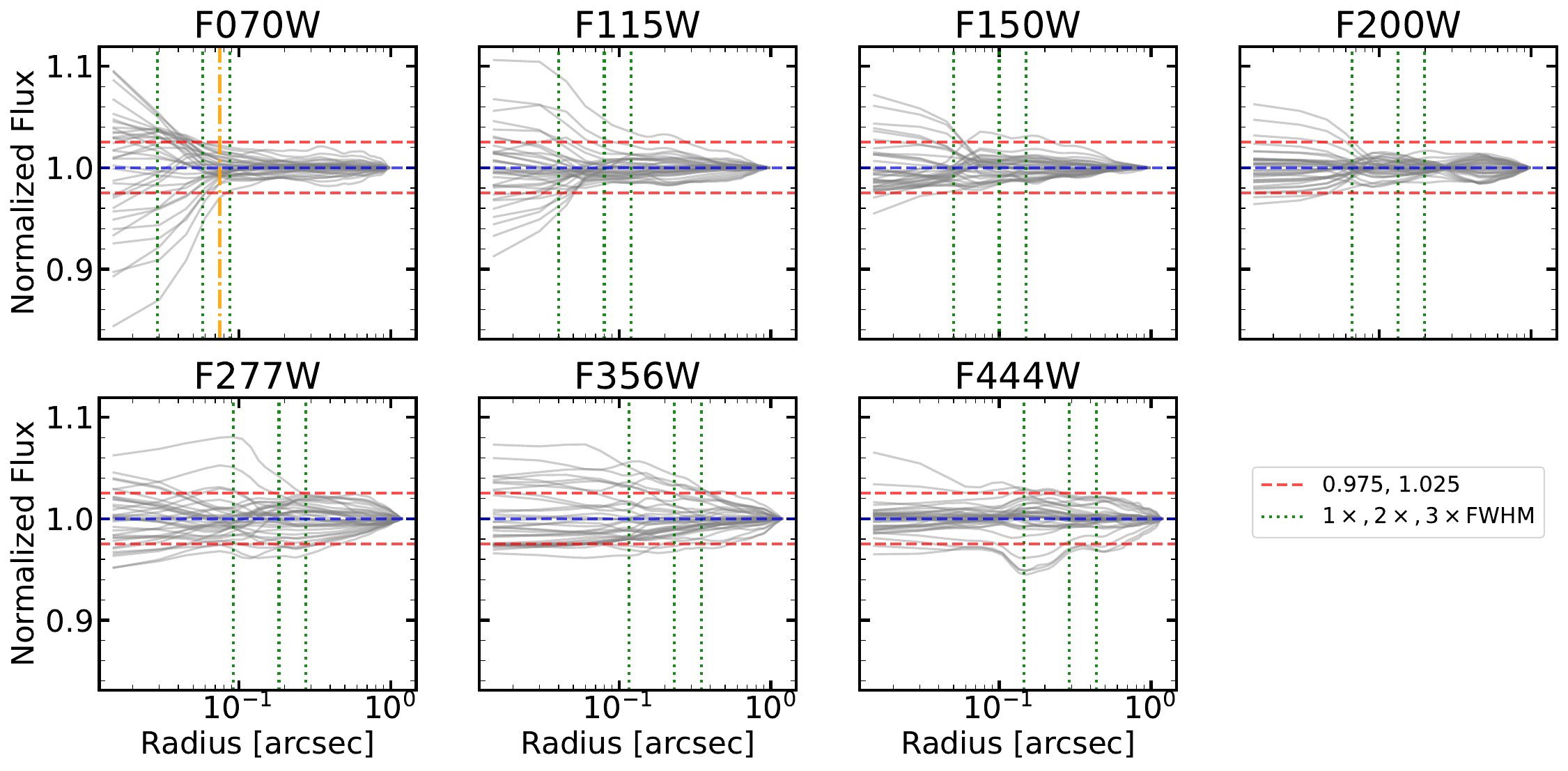}
 \centering
 \caption{Normalized curves of growth of the spatially dependent PSF models from \citet{2024ApJ...962..139Z}. The gray lines are the curves of growth of PSF models in different regions normalized by their mean curve of growth. The PSF models are normalized at $1''$ for the SW bands and 1\farcs{2} for the LW bands. The vertical green dotted lines show the radii corresponding to $1\times,~2\times$, and $3\times$ FWHM. The vertical yellow dash-dotted line for the F070W band is the radius corresponding to 2.5 pixels used for photometry. The horizontal red dashed lines correspond to 0.975 and 1.025. } 
 \label{fig:psf_vari_cof}
\end{figure*}

Here we assess how the PSF spatial variation may affect our aperture photometry. \citet{2024ApJ...962..139Z} divided the field of NIRCam into 36 ($6\times 6$) regions and constructed reliable PSF models for each region using \texttt{PSFEX}. Using their spatial dependent PSF models for the F070W, F115W, F150W, F200W, F277W, F356W, and F444W bands, we measure the curves of growth for the PSF models in different regions. We normalize the curves of growth at $1''$ for the SW bands and at 1\farcs{2} for the LW bands, where the curves of growth have flattened enough. Figure \ref{fig:psf_vari_cof} shows the curves of growth divided by their mean curves for each band. It is obvious that at a small radius, the effect of PSF variation is significant, and it mitigates at larger radii. At the radius corresponding to our photometric aperture (2.5 pixels for F070W and $2\times$FWHM for other bands), the largest variation of the curves of growth caused by the PSF spatial variation is all $\sim 5\%$ (corresponds to 0.05 mag) for each band. Therefore, we can neglect the effects caused by the PSF spatial variation. We also try to use a larger radius (2.5 pixels for F070W and $2.5\times$FWHM for each band) and find that it does not significantly affect our result of the ${\rm SNR_{\rm var}}$ distributions in Figure \ref{fig:bands_SN_distribution} and Figure \ref{fig:all_LRD_SN_distribution}. Therefore, we use $2\times$FWHM as the photometric radius, which mitigates the influence of the PSF spatial variation and achieves relatively small measurement uncertainties.

\subsection{Constraint on the LRD Variability Amplitudes}
\label{subsec:constraint_amplitude}

We have shown that the overall LRD sample and the LRD sample with broad emission lines do not exhibit significant variabilities on average. The optical variability amplitude of AGN depends on the rest-frame wavelength, luminosity, Eddington ratio, BH mass, etc \citep[e.g.,][]{2004ApJ...601..692V,2011ApJ...728...26M,2012ApJ...758..104Z}. The host contamination also dilutes the variability of the central AGN. The LRDs and known AGNs in low redshift occupy different parameter spaces and it is difficult to make a direct comparison. In this section, we try to set constraints on the LRD variability amplitudes based on the damped random walk (DRW) model and compare them with the expectation of AGNs.

The AGN UV-optical continuum variability can be well described by the DRW model \citep{2009ApJ...698..895K,2010ApJ...708..927K,2010ApJ...721.1014M,2011ApJ...735...80Z} that can recover the structure functions of AGNs \citep[e.g.,][]{2011ApJ...728...26M}. The structure function represents the root-mean-square (rms) of the magnitude differences $\Delta m$ of an AGN sample in a given time interval $\Delta t$, i.e., a typical variability amplitude at $\Delta t$. We set constraints on the asymptotic variability ${\rm SF}_{\infty}(4000\AA)$ of the LRDs adopting a DRW model likelihood \citep[][see Appendix for details]{2024arXiv240704777K}.

Figure \ref{fig:vari_constrain}(a) shows ${\rm SF}_{\infty}(4000\AA)$ versus $M_{\rm BH}$ for the 27 LRDs with measured $M_{\rm BH}$. Five LRDs were also reported by \citet{2024arXiv240704777K}. Compared to their result, our constraints are looser, because we have calibrated the photometric errors and most of the errors become larger (see Section \ref{subsec:correction_phot_uncertainty} and Figure \ref{fig:correction_curves}).
For comparison, in the figure we also plot the empirical models from \citet{2023MNRAS.518.1880B} with a varying host contamination factor $f_{*}$,\footnote{The observed host diluted rms variability amplitude is represented as:${\rm SF}_{\infty}' = \frac{L_{\rm AGN}}{L_{\rm AGN} + f_{*}L_{*}}{\rm SF}_{\infty}$, where $L_{\rm AGN}$, $L_{\rm *}$ and ${\rm SF}_{\infty}$ are the mean AGN luminosity, the host galaxy luminosity in a given band, and the intrinsic rms variability amplitude of AGN, respectively. The contamination factor $f_{*}$ (i.e., covering factor) accounts for the aperture effects, defined as the ratio between the host galaxy luminosity enclosed in an aperture and the total luminosity from the host galaxy starlight} the observed values of ${\rm SF}_{\infty}(4000\AA)$ for the Sloan Digital Sky Survey (SDSS) Stripe82 quasars \citep{2010ApJ...721.1014M}, and the Zwicky Transient Facility (ZTF) dwarf AGNs at $z \approx 0.03$ \citep{2023MNRAS.518.1880B}. The \citet{2023MNRAS.518.1880B} model assumes an Eddington ratio of 0.1, the $M_{\rm BH}\textrm{--}M_{*}$ relation from \citet{2015ApJ...813...82R}, and the stellar mass-to-light ratios of host galaxy color index of $g - r = 0.5~\rm mag$.

For a significant number of the LRDs in Figure \ref{fig:vari_constrain}, the upper limits of ${\rm SF}_{\infty}(4000\AA)$ indicate that their variability amplitudes are smaller than that expected for AGNs. The variability amplitudes for some LRDs are even smaller than the \citet{2023MNRAS.518.1880B} model with the maximum host dilution ($f_* =100\%$). We note that an Eddington ratio of 0.1 represent a model with the highest expected variability for dwarf AGNs. This is because a higher Eddington ratio leads to a lower intrinsic variability, while a lower Eddington ratio is associated with a larger host contamination \citep{2023MNRAS.518.1880B} that reduces the variability amplitude. Since most LRDs are shown to host overmassive SMBHs compared to the local $M_{\rm BH}\textrm{--}M_{*}$ relation \citep[e.g.,][]{2023ApJ...959...39H,2023ApJ...953L..29L,2024arXiv241104446C,2024A&A...691A.145M}, their host contamination is expected to be smaller than typical AGNs or local dwarf AGNs. Therefore, the non-detection of variability suggests that they are intrinsically not strongly variable sources.

There are also some LRDs with ${\rm SF}_{\infty}(4000\AA)$ upper limits $>-1$. This does not necessarily mean that they are variable, because the upper limit here is a conservative value due to our assumption above and the sparse sampling of the observations. The histogram of the ${\rm SF}_{\infty}(4000\AA)$ upper limits for all 314 LRDs in Figure \ref{fig:vari_constrain}(b) shows that sources with more multi-epoch observations (e.g., $>20$ observations in all bands) tend to have lower ${\rm SF}_{\infty}(4000\AA)$ upper limits. We also check the 11 LRDs that have $\sigma_{\rm d}(4000 \AA)$ upper limits $>-1$ in Figure \ref{fig:vari_constrain}(a) and find that 10 of them do not show ${\rm SNR_{\rm var}}$ value of larger than 2. Therefore, most of these sources are not variable either.

\begin{figure}[t]
\hspace{-0.4cm}
 \includegraphics[width=0.495\textwidth]{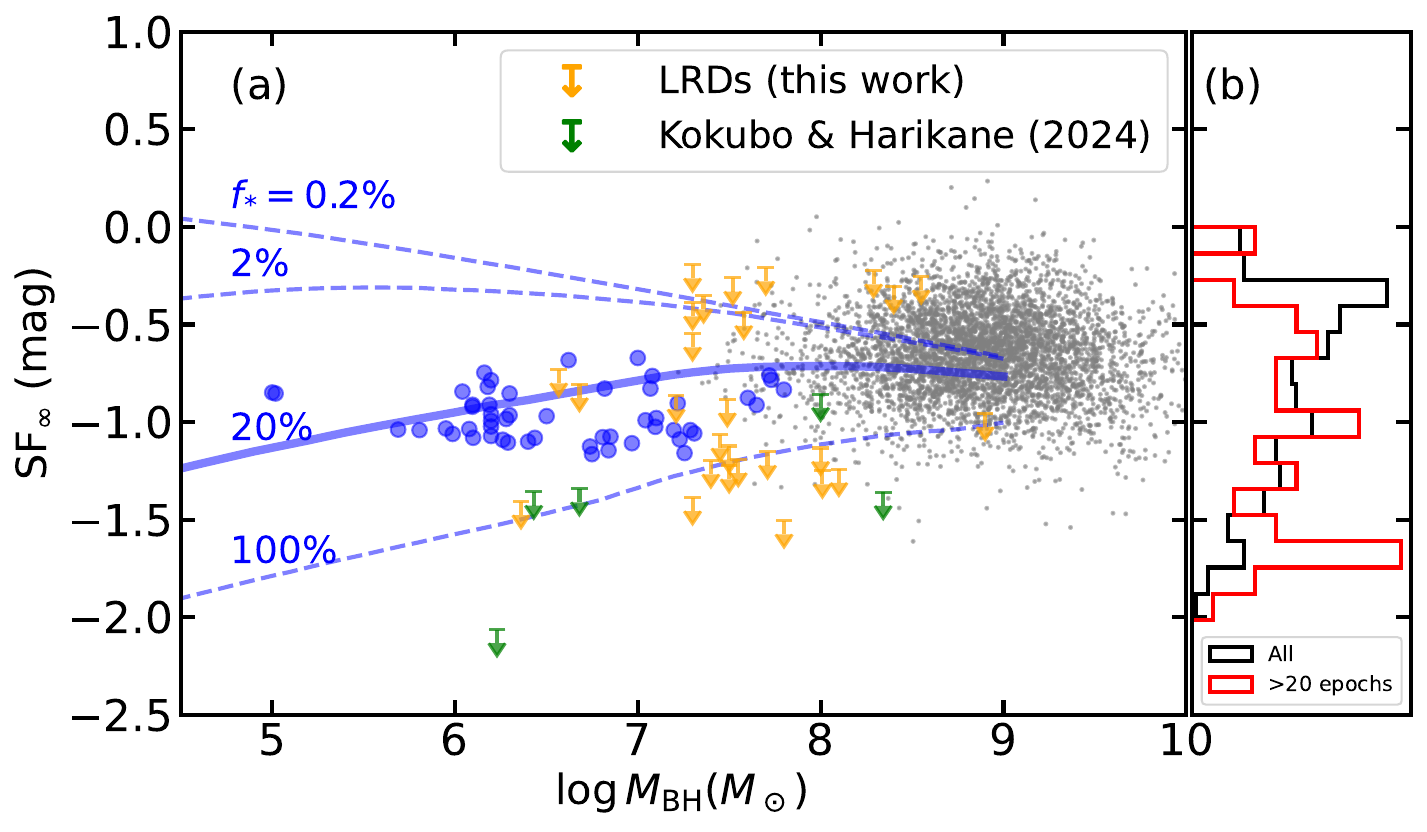}
 \centering
 \caption{(a) Asymptotic variability ${\rm SF}_{\infty}(4000\AA)$ versus $M_{\rm BH}$. The yellow downward arrows represent the upper limit of ${\rm SF}_{\infty}(4000\AA)$ at $90\%$ confidence level for the 27 LRDs with $M_{\rm BH}$ measurements. The SMBH masses are collected from the literature (Section \ref{sec:sample}). The green downward arrows represent the 5 LRDs in \citet{2024arXiv240704777K}. The blue and black points represent the ZTF dwarf AGNs from \citet{2023MNRAS.518.1880B} and the SDSS Stripe82 quasars from \citet{2010ApJ...721.1014M}, respectively. The thick solid and thin dashed lines indicate the \citet{2023MNRAS.518.1880B} empirical model of ${\rm SF}_{\infty}$ with varying host dilution covering factors around 0.2\%--100\%. (b) Distributions of ${\rm SF}_{\infty}(4000\AA)$ for all 314 LRDs that have multi-epoch observation (the black histogram) and for LRDs that have more than 20 observation across all bands (the red histogram).} 
 \label{fig:vari_constrain}
\end{figure}

\subsection{Weak LRD Variability and High AGN Accretion Rate}
\label{subsec:High_Edd_ratio}

Our LRDs have shown very weak variability on average. In addition, previous studies indicated that their X-ray emission is very weak. These two features can be explained if LRDs are dominated by normal galaxies. On the other hand, some LRDs show high EWs of broad H$\alpha$ emission and the presence of O I lines, which is consistent with the expected signatures of super-Eddington accreting BHs \citep{2022ApJ...931L..25I,2025ApJ...980L..27I}. Here we argue that the absence of strong LRD variabilities may be due to the high accretion rates of the BHs in LRDs.

Previous works found that the AGN variability amplitude decreases with the increasing Eddington ratio at a given bolometric luminosity \citep[$L_{\rm bol}$; e.g.,][]{2010ApJ...721.1014M,2012ApJ...758..104Z}. At low redshift, AGNs with very high accretion rates, such as narrow-line Seyfert 1 (NLS1) galaxies, are found to show a very low level of optical variability \citep[e.g.,][]{2004ApJ...609...69K,2016ApJ...825..126D}. The accretion rates of AGNs in LRDs are estimated to be super-Eddington or sub-Eddington \citep[e.g.,][]{2023ApJ...959...39H,2024A&A...691A.145M}, so it is possible that their lack of strong variabilities is linked to their high accretion rates. Here we simulate the expected ${\rm SNR}_{\rm var}$ distribution of LRDs to test this hypothesis.

We adopt the structure function model from \citet{2020ApJ...894...24K}, which considers the dependence on the rest-frame wavelength $\lambda_{\rm rest}$, $L_{\rm bol}$, and rest-frame time interval $\Delta t_{\rm rest}$ (see their Equations 13--22 and Table 7). These three parameters are already determined from our data, and an expected ${\rm SF}$ value can be derived for each pair. However, their structure function does not consider the dependence on the Eddington ratio. Here we incorporate the Eddington ratio dependence using Equation 9 from \citet{2023MNRAS.518.1880B}:
\begin{equation}
\label{eq2}
\begin{aligned}
    &\log \left({\frac{\rm SF_{\infty}}{\rm mag}}\right) = -0.51 - 0.479 \log \left({\frac{\rm \lambda_{\rm rest}}{\rm 4000~\AA}}\right) \\
    &+ 0.131(M_i + 23) + 0.18 \log \left({\frac{M_{\rm BH}}{10^9~\rm M_\odot}}\right), 
\end{aligned}
\end{equation}
and substitute $\frac{M_{\rm BH}}{\rm M_{\odot}} =L_{\rm bol}/\lambda_{\rm Edd}/(1.28\times10^{38}~\rm erg~s^{-1})$. We first solve Equation \ref{eq2} to derive the effective Eddington ratio for each pair using the ${\rm SF}_{\infty}(4000\AA)$ value provided by the structure function model in \citet{2020ApJ...894...24K} and the $\tau_{\rm d}$ value (see Appendix) calculated using the Equation 10 in \citet{2023MNRAS.518.1880B}. We then apply Equation \ref{eq2} to scale the SF value to the target Eddington ratio. We have assumed that the $\lambda_{\rm Edd}$ dependence in Equation \ref{eq2} can be extrapolated to a high $\lambda_{\rm Edd}$ regime. For each pair, the expected observed $\Delta m$ value is the intrinsic variability randomly drawn from a Gaussian distribution $\mathcal{N}(0, \text{SF}^2)$ plus a measurement uncertainty randomly drawn from $\mathcal{N}(0, \Delta m_{\rm err}^2)$. The simulated ${\rm SNR}_{\rm var}$ value is the ratio of the expected $\Delta m$ and the real measured error $\Delta m_{\rm err}$.

\begin{figure}[t]
\hspace{-0.4cm}
 \includegraphics[width=0.465\textwidth]{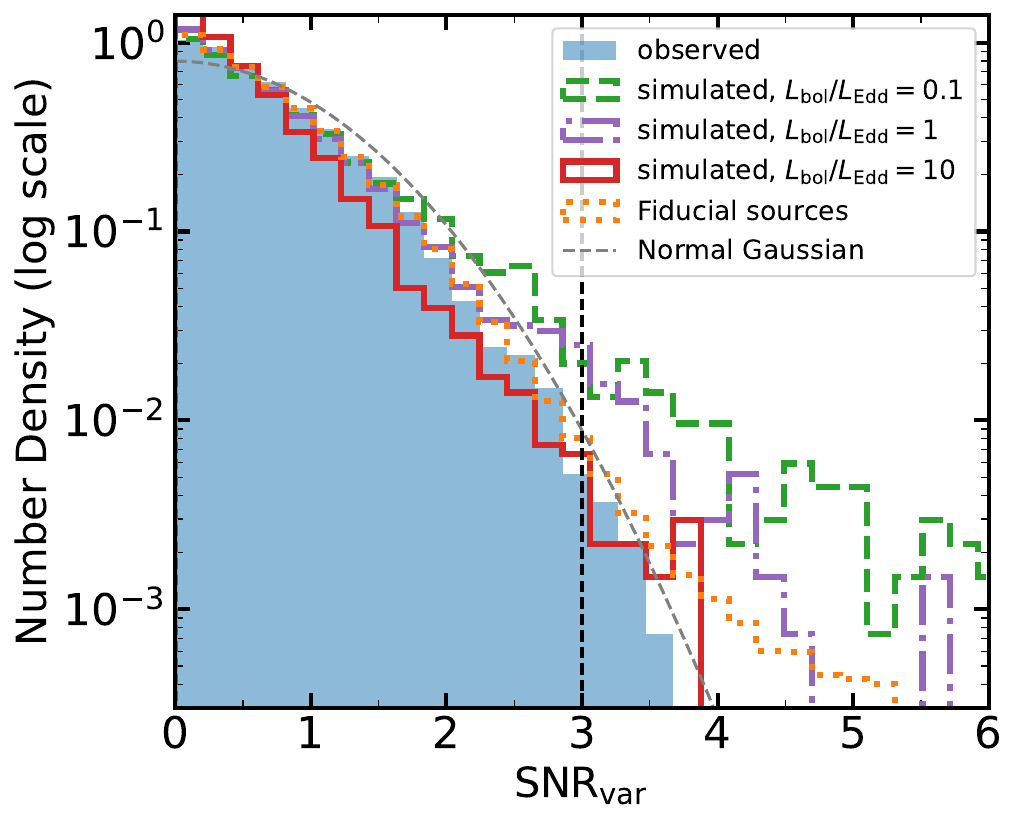}
 \centering
 \caption{Observed and simulated ${\rm SNR}_{\rm var}$ distributions of the LRD sample for all NIRCam data. The observed LRD distribution and fiducial source distribution are the same as those in Figure \ref{fig:all_LRD_SN_distribution}(g). The green dashed line, purple dash-dotted line, and red solid line represent the simulated LRD distributions with $L_{\rm bol}/L_{\rm Edd}=0.1,~1,~10$, respectively} 
 \label{fig:simulated_LRD_SNR}
\end{figure}

We compare the simulated and observed ${\rm SNR}_{\rm var}$ distributions of the LRDs in Figure \ref{fig:simulated_LRD_SNR}. For simplicity, our simulated distributions assume that all LRDs have the same Eddington ratio of 0.1, 1, and 10, respectively. In the relatively low Eddington ratio case ($\lambda_{\rm Edd}=0.1$), we expect a broader ${\rm SNR}_{\rm var}$ distribution than the observed one. However, the observed distribution well matches the simulated distribution with a super-Eddington accretion ($\lambda_{\rm Edd}=10$).

Both analytical studies and radiation hydrodynamic simulations suggest that BHs with very high accretion rates do not necessarily produce proportionally high radiative luminosities \citep[e.g.,][]{2000PASJ...52..133W,2009ApJS..183..171S,2014ApJ...796..106J}. BHs with accretion rates of $\dot{M}/\dot{M}_{\rm Edd}=10^1$--$10^3$ exhibit radiative luminosities only within the range of 1--$10~L_{\rm Edd}$ \citep[e.g., Figure 5 of][]{2020ARA&A..58...27I}, so LRDs around the super-Eddington threshold may still have very high intrinsic accretion rates. Therefore, this super-Eddington accretion scenario potentially explains the low variability amplitude of LRDs. In addition, this scenario naturally accounts for the X-ray weakness of LRDs \citep[e.g.,][]{2024arXiv241203653I,2024ApJ...976L..24M}.

\section{Conclusion}
\label{sec:conclusion}

In this study, we conduct a comprehensive investigation of the photometric variability of LRDs based on all publicly available multi-wavelength multi-epoch NIRCam data in five JWST deep fields: UDS, GOODS-S, GOODS-N, Abell 2744, and COSMOS. We also incorporate available multi-epoch JWST/MIRI data in the GOODS-S field. We compile a large sample of 806 LRDs in these five fields from the literature and find 314 of them have multi-epoch detections. With the correction of systematic photometric zero point offsets and calibration of photometric uncertainties, we evaluate the variability of these LRDs using the distribution of the signal-to-noise ratio of their variability (${\rm SNR}_{\rm var}$).

The ${\rm SNR}_{\rm var}$ distribution of LRDs is consistent with the fiducial distribution and the standard Gaussian distribution across the five fields, which indicates that these 314 LRDs on average do not show significant photometric variability. This result persists even for the 27 LRDs that exhibit broad H$\alpha$/H$\beta$ lines, which are considered to be the most reliable candidates of AGNs within the LRD sample. The estimated conservative upper limits on the DRW asymptotic variability amplitude for quite a few LRDs are inconsistent with that observed in known AGNs, even if significant host galaxy contamination is considered. These results indicate that LRDs show very weak intrinsic variability that is likely caused by the super-Eddington accretion of the BHs. Alternatively, LRDs are dominated by pure galaxies in the observed bands. We test the former scenario and find that the observed ${\rm SNR}_{\rm var}$ distribution of LRDs can be well simulated with the assumption that LRDs have very high accretion rates.

Despite the overall lack of strong variability, we find eight LRDs with strong variability of ${\rm SNR}_{\rm var}>3$, with variability amplitudes ranging from 0.24 to 0.82 mag. These objects stand out as reliable candidates that show AGN activity, though some of them could be false detections. We confirm that the variability of two of them (COS-584 and COS-593) are robust detections because of their relatively large variability amplitudes and coordinated trend of variability in different bands.

Our results demonstrate that the multi-epoch observations of JWST can be used to study the variability of a large sample of LRDs and offer new insights into their nature. Our current analysis is limited by the sparse sampling of the observations and the short time baseline. This can be improved by future JWST programs. For example, the upcoming COSMOS-3D Survey \citep{2024jwst.prop.5893K} will perform new observations in the F115W band in the COSMOS field. The combination of this program and the COSMOS-Web Survey \citep{2023ApJ...954...31C} can provide multi-epoch, F115W-band observations for most LRDs in COSMOS with a long temporal coverage. The multi-epoch JWST imaging/spectroscopic data from the upcoming North ecliptic pole EXtragalactic Unified Survey \citep[NEXUS;][]{2024arXiv240812713S} can provide multi-epoch observations with high cadence sampling and a long temporal coverage for LRDs in this field. Moreover, the rich multi-epoch data from JWST can independently identify high-redshift faint AGNs that are missed by other methods through variability, as demonstrated by \citet{2024ApJ...971L..16H} using multi-epoch HST data.

\begin{acknowledgments}

We acknowledge support from the National Key R\&D Program of China (2022YFF0503401), the National Science Foundation of China (12225301,11991052, 12233001), and the China Manned Space Project (CMS-CSST-2021-A04, CMS-CSST-2021-A05, CMS-CSST-2021-A06). We are grateful to the teams of the various projects for designing and preparing their JWST NIRCam/MIRI observations that enabled our search. We thank Kohei Inayoshi, Danyang Jiang, Fengwu Sun, Wei Leong Tee, and Feige Wang for helpful discussions. We thank Mingyang Zhuang for sharing the spatial dependent PSF models of NIRCam in their paper. 

\end{acknowledgments}

\facilities{JWST (NIRCam, MIRI)}

\software{\texttt{astropy} \citep{2013A&A...558A..33A,2018AJ....156..123A}, 
          \texttt{CIGALE} \citep{2005MNRAS.360.1413B,2019A&A...622A.103B},
          \texttt{dynesty} \citep{2020MNRAS.493.3132S,sergey_koposov_2024_12537467},
          \texttt{GalfitM}
          \citep{2013MNRAS.430..330H,2013MNRAS.435..623V},
          \texttt{matplotlib} \citep{Hunter2007}, 
          \texttt{numpy} \citep{Harris2020}, 
          \texttt{SExtractor} \citep{1996A&AS..117..393B},
          \texttt{TOPCAT} \citep{2005ASPC..347...29T},
          \texttt{WebbPSF}
         \citep{2014SPIE.9143E..3XP}
          }

\appendix
\restartappendixnumbering

\section{Estimation of the upper limit of asymptotic variability ${\rm SF}_{\infty}(4000\AA)$}

The DRW model characterizes quasar light curves as a stochastic process with an exponential covariance function $S(\Delta t) = \sigma_{\rm d}^2\exp{(-\left|\Delta t/\tau_{\rm d} \right|)}$, where $\sigma_{\rm d}$ is an amplitude parameter and $\tau_{\rm d}$ is a decorrelation time scale. The amplitude parameter is related to the structure function as ${\rm SF}_{\infty}(\lambda) = \sqrt{2} \sigma_{\rm d}(\lambda)$, where ${\rm SF}_{\infty} = {\rm SF}(\Delta t > \tau_{\rm d})$ is the asymptotic variability amplitude that represents the asymptotic value of the structure function when the time lag between observations $\Delta t$ exceeds the timescale $\tau_{\rm d}$ \citep{2010ApJ...721.1014M}. The variability amplitude is usually modeled by a \hbox{wavelength-dependent} power-law \citep[e.g.,][]{2010ApJ...721.1014M,2020ApJ...894...24K}:
\begin{equation}
    \sigma_{\rm d}(\lambda) = \sigma_{\rm d}(4000 \AA) \left( \frac{\lambda}{4000\AA}\right)^{\alpha_{\rm var}}.
\end{equation}

We adopt the light curve data likelihood of the DRW model described in the Section 3.2 and Appendix of \citet{2024arXiv240704777K}, which is expressed by a multivariate Gaussian, to set constraints on the variability amplitude of LRDs. The current time sampling of the data is limited (see Figure \ref{fig:restframe_wav_time}), so the decorrelation time scale cannot be well constrained. We use the empirical relation from \citet{2023MNRAS.518.1880B} to set a fixed value of $\tau_{\rm d}$ for each LRD as done by \citet{2024arXiv240704777K}. For those LRDs without BH mass measurement, we assume $\tau_{\rm d} = 70$ days, which is calculated using the median BH mass of LRDs. We find that in most cases, the degeneracy between $\sigma_{\rm d}(4000 \AA)$ and $\tau_{\rm d}$ is weak. Thus, the effect of this assumption is small. We also assume that the variability amplitude is independent of rest-wavelength. In this way, we can get a conservative upper limit on $\sigma_{\rm d}(4000 \AA)$. Therefore, the final likelihood has only one free parameter: $\sigma_{\rm d}(4000 \AA)$. We estimate its posterior for each source using the \emph{dynesty} package \citep{2020MNRAS.493.3132S,sergey_koposov_2024_12537467}, which is based on the advanced Dynamic Nested Sampling algorithm \citep{2019S&C....29..891H}. The prior of $\log \sigma_{\rm d}(4000\AA)$ is set as a uniform distribution between $-4$ and $0$. We calculate the 90\% confidence level upper limit of $\log \sigma_{\rm d}(4000\AA)$ from the estimated posterior distribution of each source.

\section{Samples of known AGNs in each field}
Table \ref{tab:known_AGNs} summarizes the known AGN samples used in this work.

\begin{table*}[h]
\begin{center}
   \caption{Samples of known AGNs that are removed in Section \ref{subsec:correction_phot_uncertainty}.\label{tab:known_AGNs}}
   \begin{tabular}{ll}
\hline\hline
Fields                 & Known AGN Sample         \\[4pt]
\hline
UDS      
    &X-ray sources classified as AGNs in \citet{2018MNRAS.478.2132C}\\
    &X-ray sources classified as AGNs in \citet{2018ApJS..236...48K}   \\
        
\hline
GOODS-S  & all sources in \citet{2022ApJ...941..191L} and \citet{2024ApJ...966..229L} that meet the AGN criteria in at least one band \\
&or via variability\\
\hline
GOODS-N  & X-ray sources classified as AGN in \citet{2016ApJS..224...15X}\\
& VLBI detected radio AGNs in \citet{2018Radcliffe}\\

\hline
COSMOS  & X-ray sources spectroscopically identified as AGNs or with $L_{\rm 2{\text{--}}10keV} >  10^{42} \rm ~erg\,s^{-1}$ in \citet{2010ApJ...716..348B}  \\

& X-ray sources spectroscopically identified as AGNs or with $L_{\rm 2{\text{--}}10keV} > 10^{42} \rm ~erg\,s^{-1}$ in \citet{2016ApJ...817...34M}  \\

& 692 AGN-host systems detected both
    in the X-ray and FIR from \citet{2017AA...602A.123L} \\
&AGNs selected by optically variability in \citet{2019AA...627A..33D} \\

\hline
   \end{tabular}
\end{center}
\end{table*}

\section{PSF spatial and temporal variations for the epoch pairs where LRD variabilities are detected}
\label{appdx:psf_vari}

We examine the effects of PSF spatial and temporal variations of the epoch pairs listed in Table \ref{tab:vari_LRDs}. For each epoch pair, we select point sources based on the criteria outlined in Section \ref{subsec:correction_phot_uncertainty} and require that they are brighter than 25 mag to ensure high SNRs. We consider sources within 1$’$ of the LRD so that they reflect the PSF variation at the LRD’s detector location. The magnitude difference $\Delta m$ of these point sources are measured using the aperture radius in Table \ref{tab:sext_params}. Their fractional flux changes are calculated as $1 - 10^{-0.4\times|\Delta m|}$.

The comparison of the fractional flux changes between LRDs and nearby point sources is shown in Figure \ref{appdx:psf_vari}. The fractional flux changes of nearby point sources are generally smaller than 5\% in each epoch pair, consistent with the results in Section \ref{subsec:effect_of_spatial_variation}. In contrast, the fractional flux changes for the LRDs are significantly larger. Since the LRDs and compared point sources are located in similar regions of the detector for each epoch, and in different regions between epochs, we confirm that the effects of PSF spatial and temporal variations can be neglected.

\begin{figure*}
\hspace{-0.4cm}
 \includegraphics[width=0.85\textwidth]{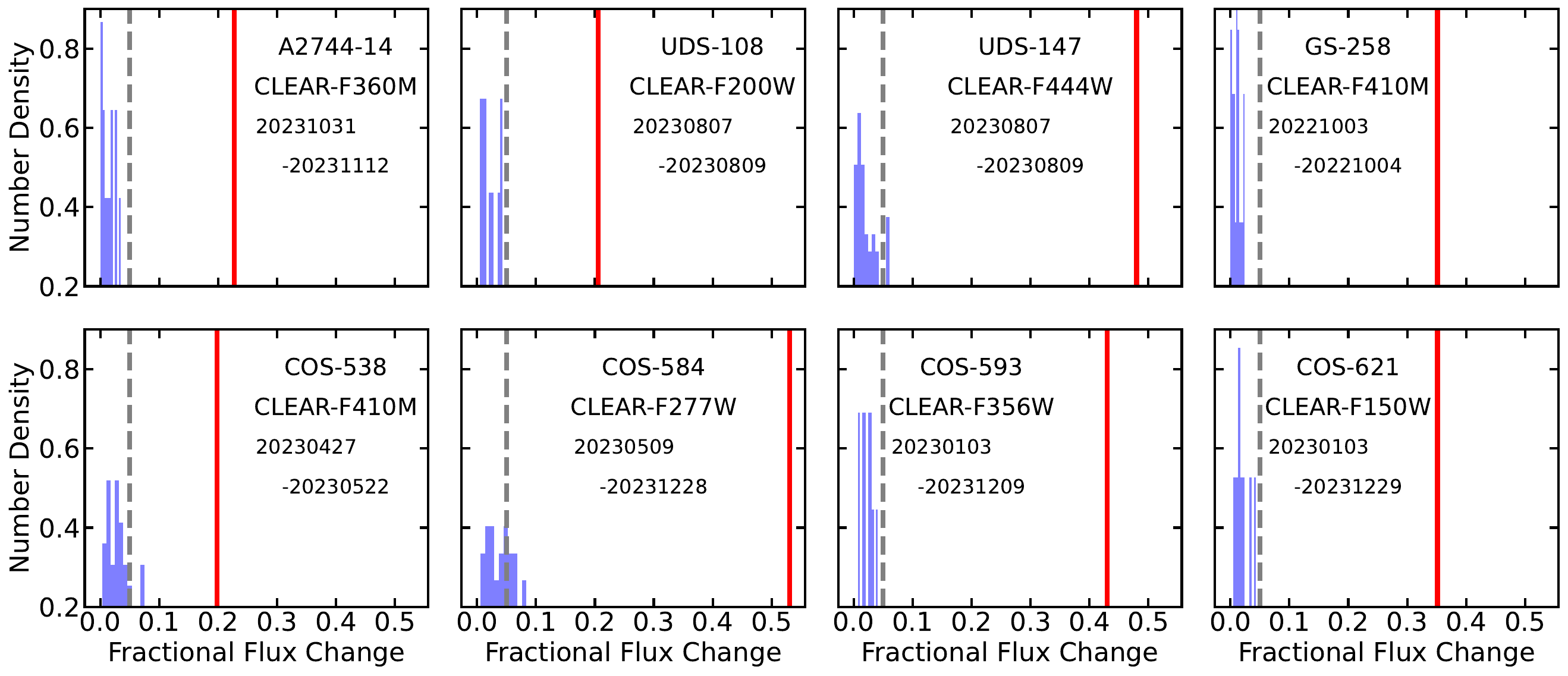}
 \centering
 \caption{Comparison of the fractional flux changes between LRDs and nearby point sources in the epoch pairs where LRD variability are detected. Each panel corresponds to one variable LRD candidate, with the source ID, filter, and date pair labeled. In each panel, the blue shaded histogram is the distribution of the fractional flux change for bright point sources ($m>25~\rm mag$) within 1$’$ of the LRD, while the vertical red solid line is the fractional flux change of LRD. The vertical gray dashed lines correspond to a fractional flux change of 0.05.}
 \label{fig:psf_vari_hist}
\end{figure*}

\section{Multi-band image fitting of the two variable LRD candidates}
\label{appxsec:galfitm}

To derive AGN fraction constraints for COS-584 and COS-593 from their morphology, we fit their multi-band JWST images using \texttt{GalfitM} \citep{2013MNRAS.430..330H,2013MNRAS.435..623V}, a multiband version of the two-dimensional image fitting code \texttt{Galfit} \citep{2002AJ....124..266P,2010AJ....139.2097P}. 

We first cut $1\farcs5 \times 1\farcs5$ cutout images for each band from the background-subtracted mosaic images. The corresponding error cutouts from the JWST pipeline are used as input sigma images. Each sigma image is scaled by a factor of $\sim 0.5$--$0.7$ to ensure that their median background pixel value is equal to the standard deviation of the background pixel values of the corresponding science image. The PSF models are constructed from the mosaic image of COSMOS field with \texttt{PSFEx} \citep{2011ASPC..442..435B}, following the method in \citet{2024ApJ...962..139Z}. We fit the eight JWST bands with a central point-source component and a \sersic\ component simultaneously. We fit the $0\farcs9 \times 0\farcs9$ region center on the source. In the fitting, the axis ratio and position angle are held constant with the wavelength, while the half-light radius and the \sersic\ index of the \sersic\ component are allowed to vary following a 2-order Chebyshev polynomial. The magnitudes of the two components are free to vary with wavelength. 

\begin{figure*}
\hspace{-0.4cm}
 \includegraphics[width=0.75\textwidth]{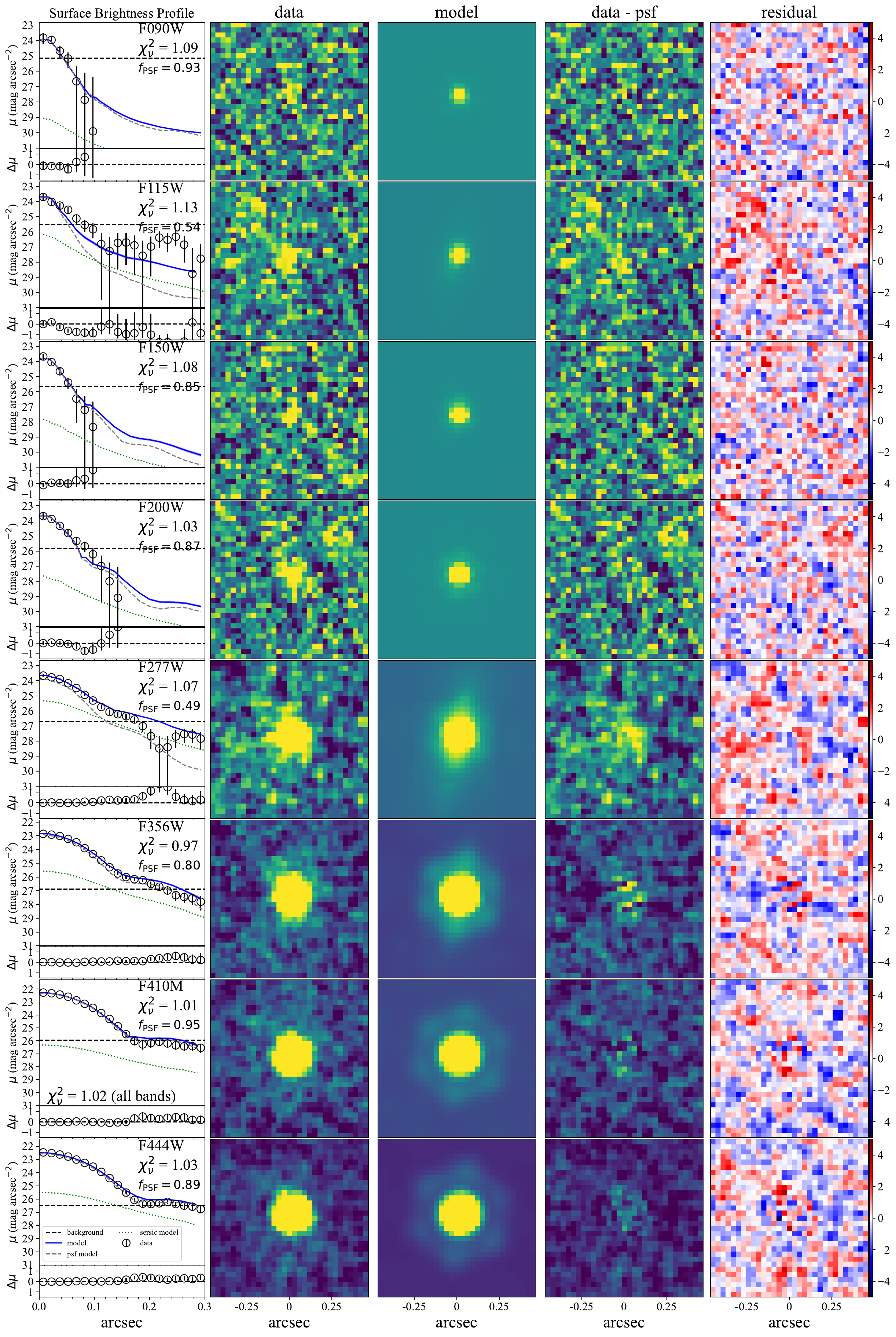}
 \centering
 \caption{Simultaneous multi-band image fitting results using a point-source model and a \sersic\ model for COS-584. Each row, from top to bottom, shows the results for the eight NIRCam bands (F090W, F115W, F150W, F200W, F277W, F356W, F410M, F444W), respectively. In the left-most column, the upper panel of each row shows the observed radial SB distribution (open circles with error bars), the PSF model (gray dashed line), the \sersic\ model (green dotted line), as well as the total model (blue solid line). The background noise level is denoted by the black horizontal dashed line. The $\chi^2$ value and point-source component fraction for each band is given in the upper-right corner of each panel. The $\chi^2$ value for all eight bands is given in the lower-left corner of the panel for F410M. The lower subpanels give the residuals between the data and the best-fit model (data$-$model). The imaging columns, from left to right, display the original data, best-fit model, data minus the nucleus point-source component, and residuals normalized by the errors (data$-$model$/$error), which are stretched linearly from $-$5 to 5.}
 \label{fig:584img_decomposion}
\end{figure*}

\begin{figure*}
\hspace{-0.4cm}
 \includegraphics[width=0.75\textwidth]{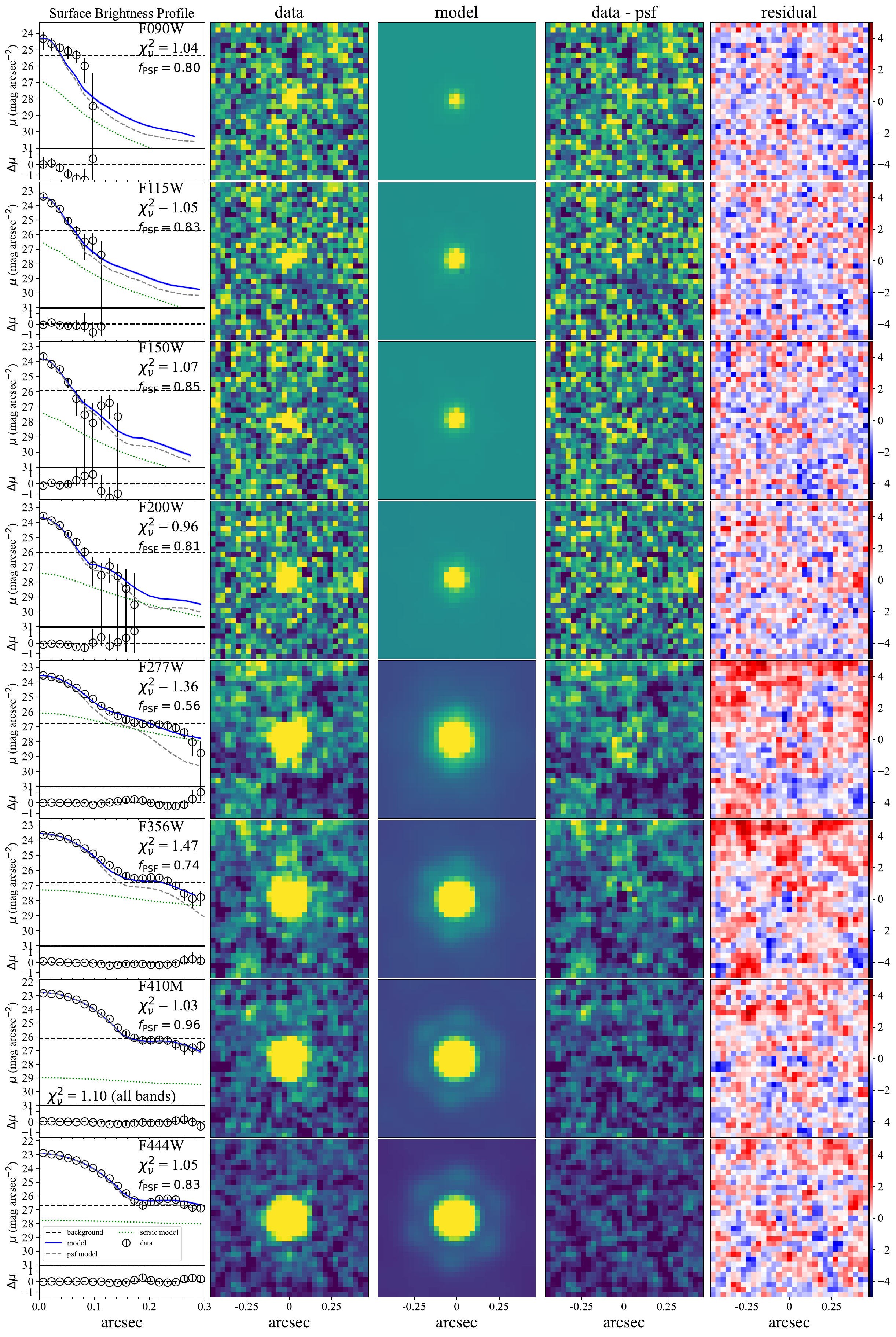}
 \centering
 \caption{Same as Figure \ref{fig:584img_decomposion}, but for COS-593.}
 \label{fig:593img_decomposion}
\end{figure*}

Figure \ref{fig:584img_decomposion} and Figure \ref{fig:593img_decomposion} show the fitting results for COS-584 and COS-593, respectively. Based on the best-fit model, we calculate the point-source component fraction ($f_{\rm psf}$) for each band, which are summarized in Table \ref{tab:agn_fraction_constraint}. The $f_{\rm psf}$ value for each band is treated as the upper limit of the AGN component during the SED fitting described in Appendix \ref{appxsec:sedfitting}.

\section{SED Fitting Setup}
\label{appxsec:sedfitting}

Table \ref{tab:cig} summarizes the adopted parameters of \texttt{CIGALE}. We consider both a stellar component and an AGN component. { A nebular emission component is included to account for the emission line, but possible broad-line contributions from the AGN are not considered in our model.} In the fitting process, we take into account the AGN fraction constraints of each band set by the morphology fitting in Appendix \ref{appxsec:galfitm} and the variability amplitude (Table \ref{tab:agn_fraction_constraint}). The constraint on the AGN fraction is achieved by modifying the code of \texttt{CIGALE} (v2025), such that any model where the AGN fraction in any band does not meet the constraint is assigned a chi-squared value of NaN.

\begin{table*}
\caption{AGN fraction constraints for each band in SED fitting}
\label{tab:agn_fraction_constraint}
\hspace*{-1.5cm}
\begin{tabular}{lccccccccc}
\hline
\hline
LID&&F090W&F115W&F150W&F200W&F277W&F356W&F410M&F444W \\
\hline
COS-584&$f_{\rm psf}~(f_{\rm AGN, upper})$&0.93&0.54&0.85&0.87&0.49&0.80&0.95&0.89 \\
&$f_{\rm AGN, lower}$ &-&-&-&-&0.36&-&-&-\\
\hline
COS-593&$f_{\rm psf}~(f_{\rm AGN, upper})$&0.80&0.83&0.85&0.81&0.56&0.74&0.96&0.83\\
&$f_{\rm AGN, lower}$&-&-&-&-&-&0.24&-&-\\

\hline
\end{tabular}
\end{table*}

\begin{table*}
\caption{\textsc{cigale} model parameters}
\hspace{-1.5cm}
\label{tab:cig}
\begin{tabular}{llll} \hline\hline
Module & Parameter & Symbol & Values \\
\hline
\multirow{2}{*}{\shortstack[l]{Star formation history\\
                               \texttt{sfhdelayed} }}
    & Stellar e-folding time & $\tau_{\rm star}$ & 0.2, 0.5, 1, 2, 3, 4, 5 Gyr\\
    & Stellar age & $t_{\rm star}$  
            & \fst{0.1, 0.2, 0.5,} 1, 2, 3, 4, 5, 7 Gyr\\ 
\hline
\multirow{2}{*}{\shortstack[l]{Simple stellar population\\ 
    \texttt{bc03} }}
    & Initial mass function & $-$ & \cite{2003ApJ...586L.133C} \\
    & Metallicity & $Z$ & 0.02 \\
\hline
\multirow{2}{*}{\shortstack[l]{Nebular emission\\
                               \texttt{nebular} }} 
    & Ionization parameter & $\log U$ & $-4.0$, $-2.0$\\
    & Gas metallicity & $Z_{\rm gas}$ & 0.0004, 0.002, 0.033 \\
    & Fraction of escaped Lyman continuum photons & $f_{\rm esc}$ & 0.0, 0.05, 0.1, 0.5, 0.7, 1.0 \\
\hline
\multirow{3}{*}{\shortstack[l]{Dust attenuation \\ 
                \texttt{dustatt\_modified\_starburst} }}
    & \multirow{2}{*}{\shortstack[l]{Color excess of nebular lines}} & \multirow{2}{*}{\shortstack[l]{$E(B-V)_{\rm line}$}} & \multirow{2}{*}{\shortstack[l]{0,0.02,0.05,0.1,0.2,\\
                              0.3,0.4,0.5,0.7,0.9,1.0}} \\\\
    & ratio between line and continuum $E(B-V)$ & $\frac{E(B-V)_{\rm line}}{E(B-V)_{\rm cont}}$ & 0.44, 0.8, 1\\
\hline
\multirow{4}{*}{\shortstack[l]{Galactic dust emission \\ \texttt{dl2014} }}
    & PAH mass fraction & $q_{\rm PAH}$ & 0.47, 2.5, 7.32 \\
    & Minimum radiation field & $U_{\rm min}$ & 0.1, 1.0, 10, 50 \\
    & \multirow{2}{*}{\shortstack[l]{Fraction of PDR emission}} & \multirow{2}{*}{\shortstack[l]{$\gamma$}} & \multirow{2}{*}{\shortstack[l]{0.01, 0.02, 0.05, \\
                                            0.1, 0.2, 0.5, 0.9}} \\\\
\hline
\multirow{5}{*}{\shortstack[l]{AGN (UV-to-IR) emission \\ \texttt{skirtor2016} }}
    & Average edge-on optical depth at $9.7 \mu$m & $\tau_{9.7}$ & 3,5,7,9,11 \\
    & Viewing angle & $\theta_{\rm AGN}$ & 30$^\circ$, 70$^\circ$  \\
    & \multirow{2}{*}{\shortstack[l]{AGN contribution at given wavelength}} & \multirow{2}{*}{\shortstack[l]{$\fracA$}} & \multirow{2}{*}{\shortstack[l]{0.1,0.2,0.3,0.4,0.5,\\0.6,0.7,0.8,0.9,0.95,0.99}}\\\\
    & Wavelength range where $\fracA$ is defined & $\lambda_{\rm AGN}$ &  {\shortstack[l]{0/0}}\\\\
\hline
    \multirow{2}{*}{\shortstack[l]{Redshift$+$IGM \\ \texttt{redshifting} }} 
    &  \multirow{2}{*}{\shortstack[l]{ Source redshift }} & \multirow{2}{*}{\shortstack[l]{ $z$ }} & \multirow{2}{*}{\shortstack[l]fixed as phot-$z$s} \\\\
\hline
\end{tabular}
\end{table*}


\end{document}